\documentclass[preprint, tighten, twocolumn]{aastex62}

\newcommand{\swift}{{\it Swift}}
\newcommand{\hst}{{\it HST}}
\newcommand{\nga}{NGC\, 5548}
\newcommand{\ngb}{NGC\, 4151}
\newcommand{\ngc}{NGC\, 4593}
\newcommand{\mrk}{Mrk\, 509}
\newcommand{\et}{et al.\,}
\newcommand{\javelin}{{\tt JAVELIN}}
\newcommand{\rb}{}
\shorttitle{Swift AGN disk RM Survey}
\shortauthors{Edelson \et}

\begin{document}

\title{The First Swift \rb{Intensive} AGN Accretion Disk Reverberation Mapping Survey}

\author[0000-0001-8598-1482]{R. Edelson}
\affiliation{Department of Astronomy, University of Maryland, College Park, MD 20742-2421, USA}
\email{email: redelson@astro.umd.edu}

\author[0000-0001-9092-8619]{J. Gelbord}
\affiliation{Spectral Sciences Inc., 4 Fourth Ave., Burlington, MA 01803, USA}

\author[0000-0002-8294-9281]{E. Cackett}
\affiliation{Department of Physics and Astronomy, Wayne State University, 666 W. Hancock St, Detroit, MI 48201, USA}

\author[0000-0001-6481-5397]{B.M. Peterson}
\affiliation{Department of Astronomy; The Ohio State University; 140 West 18th Ave.; Columbus, OH 43210, USA}
\affiliation{Center for Cosmology and AstroParticle Physics; The Ohio State University; 192 West Woodruff Ave., Columbus, OH 43210, USA}
\affiliation{Space Telescope Science Institute; 3700 San Martin Drive; Baltimore, MD 21218, USA}

\author[0000-0003-1728-0304]{K. Horne}
\affiliation{SUPA Physics and Astronomy, University of St. Andrews, Fife, KY16 9SS, Scotland, UK}

\author[0000-0002-3026-0562]{A. J. Barth}
\affiliation{Department of Physics and Astronomy, 4129 Frederick Reines Hall, University of California, Irvine, CA, 92697-4575, USA}

\author[0000-0002-6408-1001]{D.A. Starkey}
\affiliation{SUPA Physics and Astronomy, University of St. Andrews, Fife, KY16 9SS, Scotland, UK}

\author[0000-0002-2816-5398]{M. Bentz}
\affiliation{Department of Physics and Astronomy, Georgia State University, Atlanta, GA 30303, USA}

\author[0000-0002-0167-2453]{W.N. Brandt}
\affiliation{Department of Astronomy and Astrophysics, 525 Davey Lab, The Pennsylvania State University, University Park, PA 16802, USA}
\affiliation{Institute for Gravitation and the Cosmos, The Pennsylvania State University, University Park, PA 16802, USA}
\affiliation{Department of Physics, The Pennsylvania State University, University Park, PA 16802, USA}

\author{M. Goad}
\affiliation{University of Leicester, Department of Physics and Astronomy, Leicester, LE1 7RH, UK}

\author[0000-0003-0634-8449]{M. Joner}
\affiliation{Department of Physics and Astronomy, N283 ESC, Brigham Young University, Provo, UT 84602-4360, USA}

\author[0000-0003-0944-1008]{K. Korista}
\affiliation{Department of Physics; Western Michigan University; Kalamazoo, MI 49008-5252, USA}

\author{H. Netzer}
\affiliation{School of Physics and Astronomy, Raymond and Beverly Sackler Faculty of Exact Sciences, Tel Aviv University, Tel Aviv 69978, Israel}

\author[0000-0001-5624-2613]{K. Page}
\affiliation{University of Leicester, Department of Physics and Astronomy, Leicester, LE1 7RH, UK}

\author[0000-0001-9355-961X]{P. Uttley}
\affiliation{Anton Pannekoek Institute, University of Amsterdam, Science Park 904, 1098 XH, Amsterdam, The Netherlands}

\author[0000-0003-4808-092X]{S. Vaughan}
\affiliation{University of Leicester, Department of Physics and Astronomy, Leicester, LE1 7RH, UK}

\author[0000-0002-0001-7270]{A. Breeveld}
\affiliation{Mullard Space Science Laboratory, University College London, Holmbury St. Mary, Dorking, Surrey. RH5 6NT, UK}

\author[0000-0003-1673-970X]{S. B. Cenko}
\affiliation{Department of Astronomy, University of Maryland, College Park, MD 20742-2421, USA}
\affiliation{Astrophysics Science Division, NASA Goddard Space Flight Center, Mail Code 661, Greenbelt, MD 20771, USA}
\affiliation{Joint Space-Science Institute, University of Maryland, College Park, MD 20742, USA}

\author{C. Done}
\affiliation{Centre for Extragalactic Astronomy, Department of Physics, University of Durham, South Road, Durham, DH1 3LE, UK}

\author[0000-0002-8465-3353]{\rb{P. Evans}}
\affiliation{University of Leicester, Department of Physics and Astronomy, Leicester, LE1 7RH, UK}

\author{M. Fausnaugh}
\affiliation{MIT Kavli Institute for Space and Astrophysics Research, 77 Massachusetts Avenue, 37-241, Cambridge, MA 02139, USA}

\author[0000-0003-4503-6333]{G. Ferland}
\affiliation{Department of Physics, University of Kentucky, Lexington KY 40506, USA}

\author[0000-0002-9280-1184]{D. Gonzalez-Buitrago}
\affiliation{Department of Physics and Astronomy, 4129 Frederick Reines Hall, University of California, Irvine, CA, 92697-4575, USA}

\author{J. Gropp}
\affiliation{Department of Astronomy and Astrophysics, 525 Davey Lab, The Pennsylvania State University, University Park, PA 16802, USA}

\author[0000-0002-9961-3661]{D. Grupe}
\affiliation{Department of Earth and Space Sciences, Morehead State University, 235 Martindale Dr, Morehead, KY 40351, USA}

\author[0000-0001-5540-2822]{J. Kaastra}
\affiliation{SRON Netherlands Institute for Space Research, Sorbonnelaan 2, 3584 CA Utrecht, the Netherlands}
\affiliation{Leiden Observatory, Leiden University, P.O. Box 9513, 2300 RA Leiden, the Netherlands}

\author{J. Kennea}
\affiliation{Department of Astronomy and Astrophysics, 525 Davey Lab, The Pennsylvania State University, University Park, PA 16802, USA}

\author[0000-0002-2180-8266]{G. Kriss}
\affiliation{Space Telescope Science Institute; 3700 San Martin Drive; Baltimore, MD 21218, USA}

\author{S. Mathur}
\affiliation{Department of Astronomy; The Ohio State University; 140 West 18th Ave.; Columbus, OH 43210, USA}
\affiliation{Center for Cosmology and AstroParticle Physics; The Ohio State University; 192 West Woodruff Ave., Columbus, OH 43210, USA}

\author[0000-0002-4992-4664]{M. Mehdipour}
\affiliation{SRON Netherlands Institute for Space Research, Sorbonnelaan 2,
3584 CA Utrecht, the Netherlands}

\author[0000-0003-2371-4121]{D. Mudd}
\affiliation{Department of Physics and Astronomy, 4129 Frederick Reines Hall, University of California, Irvine, CA, 92697-4575, USA}

\author[0000-0001-7084-4637]{J. Nousek}
\affiliation{Department of Astronomy and Astrophysics, 525 Davey Lab, The Pennsylvania State University, University Park, PA 16802, USA}

\author[0000-0002-2772-8160]{T. Schmidt}
\affiliation{Department of Physics and Astronomy, 4129 Frederick Reines Hall, University of California, Irvine, CA, 92697-4575, USA}

\author[0000-0001-9191-9837]{M. Vestergaard}
\affiliation{Niels Bohr Institute, University of Copenhagen, Juliane Maries Vej 30, DK-2100 Copenhagen Ø, Denmark}
\affiliation{Steward Observatory, Dept. of Astronomy, University of Arizona, 933 N. Cherry Ave, Tucson AZ 85718, USA}

\author[0000-0002-8956-6654]{C. Villforth}
\affiliation{University of Bath, Department of Physics, Claverton Down, Bath, BA27AY, UK}

\begin{abstract}
\swift\ intensive accretion disk reverberation mapping of four AGN yielded light curves sampled $\sim$200-\rb{350} times in 0.3--10 keV X-ray and six UV/optical bands. 
\rb{Uniform reduction and cross-correlation analysis of these datasets yields three main results:
(1) The X-ray/UV correlations are much weaker than those within the UV/optical, posing severe problems for the lamp-post reprocessing model in which variations in a central X-ray corona drive and power those in the surrounding accretion disk. 
(2) The UV/optical interband lags are generally consistent with $ \tau \propto \lambda^{4/3} $ as predicted by the centrally illuminated thin accretion disk model.}
While the average interband lags are somewhat larger than predicted, these results alone are not inconsistent with the thin disk model given the large systematic uncertainties involved.
(\rb{3) The one exception is the {\it U} band lags, which are on average a factor of $\sim$2.2 larger than predicted from the surrounding band data and fits.
This excess appears to be due to diffuse continuum emission from the broad-line region (BLR).}
The precise mixing of disk and BLR components cannot be determined from these data alone.
The lags in different AGN appear to scale with mass or luminosity.
We also find that there are systematic differences between the uncertainties derived by \javelin\ vs. more standard lag measurement techniques, with \javelin\ reporting smaller uncertainties by a factor of 2.5 on average.  
In order to be conservative only standard techniques were used in the analyses reported \rb{herein}.
\end{abstract}


\section{Introduction}
\label{section:intro}

Due to their vast distances the central regions of active galactic nuclei (AGN) cannot be imaged directly, so we are forced to utilize indirect methods to discern their structure and physical conditions. 
Historically, the strong multiwavelength variability of AGN has provided perhaps our strongest physical constraints.
For instance, as first noted by \cite{Lynden-Bell69}, the combination of the rapid variability and high luminosity of AGN requires that this central region have such high densities that a supermassive black hole provides the only (known) viable explanation.
The more detailed picture of an optically thick, geometrically thin accretion disk surrounding that black hole was first proposed by \cite{Shakura73} in the context of stellar-mass black holes.
\cite{Galeev79} added magnetic reconnection in a corona above the disk in order to explain the observed hard X-ray emission from AGN. 
Under these circumstances the central corona can directly illuminate and heat the outer disk  (e.g., \citealt{Frank02}), leading to the so-called  ``lamp-post/reprocessing'' model.
Note that these models are not entirely dependent on each other: it is possible that the thin disk model could be correct but that the variations are not driven by the reprocessing of radiation from a central corona.

A clear prediction of these models is that the variable X-ray emission from the corona will illuminate and heat and thus be reprocessed and seen in the UV/optical emission from the disk.
Measurement of the interband X-ray/UV temporal lag and smoothing can then be used to estimate the size and structure of the disk.
This technique, known as reverberation mapping (RM; \citealt{Blandford82}; \citealt{Peterson93}), has been used for decades in a different context to constrain the size and physical characteristics of the broad emission line region (BLR).
For the disk, the reprocessing model predicts a clear relation between the interband lag ($\tau$) and observing wavelength ($\lambda$) as the variations from the smaller, hotter inner disk are expected to precede those from the larger, cooler outer disk regions, scaling as $ \tau \propto \lambda^{4/3} $ (e.g., \citealt{Cackett07}).

There have been many attempts to search for this expected lag structure, but until recently these have yielded inconclusive results.
Efforts to implement disk RM by correlating X-ray light curves gathered with space-based observatories with optical light curves typically from ground-based observatories (e.g., \rb{\citealt{Arevalo08},} \citealt{Breedt09}) often yielded suggestions of interband lags in the expected direction, but the results were never statistically significant ($>3\sigma$).
The practical difficulties of coordinating monitoring with such dissimilar observing constraints were too great to overcome.
Likewise, comparisons between bands in ground-based optical monitoring yielded indications that the shorter wavelengths led the longer wavelengths (e.g., \citealt{Sergeev05}; \citealt{Cackett07}), but again not at a statistically significant level.
It turns out that the experiment's focus on the optical was the right track to take, but the limited wavelength range (about an octave) was insufficient to clearly observe the expected interband lags.

The 2004 launch of the {\it Neil Gehrels Swift Observatory} (\swift\ hereafter) provided a single observatory that could monitor across the X-ray, UV and optical bands at high cadence.
The focus of the original \swift\ mission \citep{Gehrels04} was on identifying and observing $\gamma$-ray bursts (GRBs), and AGN disk RM did not at first make full use of its capabilities.
This began to change with campaigns by \cite{Shappee14} and \cite{McHardy14}, which found evidence of the UV leading the optical in NGC~2617 and NGC~5548, respectively.

These campaigns set the stage for the development of the intensive disk reverberation mapping (IDRM) technique, an observing strategy that makes full use of \swift's unique ability to monitor AGN variability across the X-ray, UV, and optical at high cadence and over long durations.
\rb{IDRM observations of four AGN (\nga, \ngb, \ngc, and \mrk) have been completed as of the end of 2017.}
This paper presents a systematic reduction and analysis of the IDRM data on \rb{these} four AGN in order to survey their interband cross-correlation properties and test the standard thin accretion disk/reprocessing picture.
The paper is organized as follows.
Section~2 summarizes the observations and data reduction,  
Section~3 presents the timing analysis, 
Section~4 discusses the theoretical implications of these results,
and Section~5 gives some brief concluding remarks.

\section{Observations and data reduction}
\label{section:data}

\subsection{Observing Strategy}
\label{section:obs}

\rb{The IDRM observing technique involves three specific improvements over RM campaigns executed before the launch of \swift\ and even in the early years of the \swift\ mission:
\begin{enumerate}
\item Monitoring with all six UVOT filters, covering the UV/optical from 1928 to 5468~\AA.
\item Sampling typically 2-3 times faster (relative to the scale size set by the Schwarzschild radius) than previous campaigns.
\item Executing a total of $\sim$200-350 visits (samples) in each campaign.
\end{enumerate}
Because previous studies typically gathered $\sim$100 or \rb{fewer} samples in one or a few bands in either the UV or the optical, these changes represent improvements of at least a factor of 2 in each of these three key quantities.
This intensive blanketing in both the time and energy domains is what makes this technique much more sensitive to very short lags, particularly across the crucial UV/optical regime, as would be expected for a signal propagating through an accretion disk at the speed of light.}

\rb{This general approach of intensive monitoring was initially approved by the \swift\ director before any specific targets were selected.
After it was learned that a large \hst\ monitoring project was approved for \nga, that target was chosen as the first \swift\ IDRM target, so that simultaneous monitoring could occur with \hst\ and a large collection of ground-based observatories.
That campaign was a success, yielding the first clear indications of lags increasing with wavelength across the UV/optical (\citealt{Edelson15}, \citealt{Fausnaugh16}).
Subsequent IDRM campaigns also detected clear interband UV/optical lags in the nearby \ngb\ \citep{Edelson17} and \ngc\ \citep{McHardy18}.}
We note that other groups have analyzed \rb{these \swift\ IDRM data}, e.g., \cite{Gardner17} and \cite{Pal17}.
IDRM monitoring of a fourth AGN, \mrk, was completed in 2017 December; we report on these results for the first time in this contribution.

The four IDRM campaigns reported herein \rb{are} all designed in a broadly similar fashion, with $\sim$200-\rb{350} visits and UVOT sampling in all six filters.
(In practice fewer samples were actually obtained, due to GRB and other interruptions in the observing programs and UVOT dropouts, as discussed in \rb{the Appendix}.)
The total duration and sampling rate of each campaign were scaled roughly by the observed luminosity of the target, so the IDRM campaign on the lowest-luminosity target, \ngc, lasted $\sim$23 days, while that on the most luminous target, \mrk, lasted $\sim$9 months.
Source and campaign parameters for these four IDRM AGN are given in Table~1.

\begin{deluxetable*}{lcccccccc}
\label{table1}
\tablenum{1}
\tablecaption{IDRM AGN source and campaign parameters
\label{tab:dropouts}}
\tablecolumns{8}
\tablehead{
\colhead{(1)} & \colhead{(2)} & \colhead{(3)} & \colhead{(4)} & \colhead{(5)} & 
 \colhead{(6)} & \colhead{(7)} & \colhead{(8)} \cr
\colhead{} & \colhead{} & \colhead{} & \colhead{} & \colhead{Date Range} &
 \colhead{Duration} & \colhead{Total} & \colhead{Mean Sampling} \cr
\colhead{Object} & \colhead{Redshift} & \colhead{log$_{10}(M_{BH}/M_\odot)$} &
 \colhead{$\dot m_{\rm Edd}$} & \colhead{(MJD)} & \colhead{(days)} &
 \colhead{Visits} & \colhead{Interval (days)}}
\startdata
Mrk~509  & 0.0341 & 8.05 & 5\% & 57829.9$-$58102.5 & 272.6 & 257 & 1.065 \cr
NGC~5548 & 0.0163 & 7.72 & 5\% & 56706.0$-$56833.6 & 127.6 & 291 & 0.440 \cr
NGC~4151 & 0.0032 & 7.56 & 1\% & 57438.0$-$57507.3 &  69.3 & 322 & 0.216 \cr
NGC~4593 & 0.0083 & 6.88 & 8\% & 57582.8$-$57605.4 &  22.6 & 194 & 0.117
\enddata
\tablecomments{Column 1: Object.
Column 2: Redshift from the SIMBAD database.
Column 3: Black hole mass from the AGN Black Hole Mass Database, as described in \cite{Bentz15}.
Column 4: Eddington ratio $\dot m_{\rm Edd} =  L_\mathrm{bol} / L_\mathrm{Edd} $. 
Column 5: MJD range of the campaign.
Column 6: Campaign duration in days.
Column 7: Total number of ``good'' visits (in which usable data were gathered in at least one band).
Column 8: mean sampling interval based on the values in Columns 6 and 7.}
\end{deluxetable*}

\subsection{UVOT Data Reduction}
\label{section:uvot}

This paper's UVOT data reduction follows the same general procedure described in our previous work on \nga\ \citep{Edelson15} and \ngb\ \citep{Edelson17}.
Thus, this paper will present only a broad overview of this process; the reader is referred to these earlier papers for more detailed descriptions.
This process has three steps: flux measurement, removal of points that fail quality checks, and identification and masking of low sensitivity regions of the detector.
Each step is described in turn below.

All data were reprocessed for uniformity (using version 6.22.1 of {\tt HEASOFT})
and their astrometry refined (following the procedure of \citealp{Edelson15})
before measuring fluxes using {\tt UVOTSOURCE} from the {\tt FTOOLS}\footnote{http://heasarc.gsfc.nasa.gov/ftools/} package \citep{Blackburn95}.
The filters and other details of this instrument are given in \cite{Poole08}.
Source photometry was measured in a circular extraction region of 5\arcsec\ in radius, while backgrounds were taken from concentric 40\arcsec--90\arcsec\ annuli.
Note that the underlying galaxy can contribute to both of these regions, especially in the nearby AGN \ngc\ and \ngb.
In the {\it V} band in particular, especially when the AGN power is lower,the galaxy contributes significantly to the measured flux.
This decreases the apparent variability/noise ratio and lowers the correlation coefficients (see Section 3.2.2).
The final flux values include corrections for aperture losses, coincidence losses, large-scale variations in the detector sensitivity across the image plane, and declining sensitivity of the instrument over time.

In the second step, the resulting measurements are used for both automated quality checks and to flag individual observations for manual inspection. 
These automated checks include aperture ratio screenings to catch instances of extended point-spread functions (PSFs) or when the astrometric solution is off, 
the elimination of short, full-frame safety check exposures (taken prior to data collected in much longer hardware window exposures), and a minimum exposure time threshold of 20 s.
Data are flagged for inspection when the fitted PSFs of either the AGN or several field stars were found to be unusually large or asymmetric, or if fewer than 10 field stars with robust centroid positions are available for astrometric refinement.  
Upon inspection, observations are rejected if there were obvious astrometric errors, doubled or distorted PSFs, or prominent image artifacts (e.g., readout streaks or scattered light) that would affect the AGN measurement.
The IDRM observations consist of 6411 exposures after eliminating safety frames and 33 short exposures; from these, 53 are screened out (43 failed automated tests and 28 failed manual inspections), yielding a final set of 6358 exposures for all four targets combined.
Note that we have adopted a non-standard setting of 7.5\% for the {\tt UVOTSOURCE} parameter {\tt FWHMSIG} because this yields flux uncertainties more consistent with Gaussian statistics \citep{Edelson17}.

\rb{The third step was to use the apparent dropouts from the UVOT light curves to identify detector regions with reduced sensitivity, then to define UV and optical detector masks to screen out all points that fall within these regions.  
This process and the resulting masks are described in the Appendix.}

\subsection{XRT Data Reduction}
\label{section:xrt}

The \rb{XRT} data were analyzed using the \rb{standard \swift\ analysis} tools described by \cite{Evans09}\footnote{http://www.swift.ac.uk/user\_objects}.
\rb{These} produce light curves that are fully corrected for instrumental effects such as pile up, dead regions on the CCD, and vignetting.
\rb{The source aperture varies dynamically according to the source brightness and position on the detector. 
For full details see \cite{Evans07} and \cite{Evans09}.}

\rb{We output the observation times (the midpoint between the start and end times) in MJD instead of the default of seconds since launch for ease of comparison with the UVOT data.}
We utilize ``snapshot'' binning, which produces one bin for each continuous spacecraft pointing. 
This is done because these short visits always occur completely within one orbit with one set of corresponding exposures in the UVOT filters.
\rb{In all other cases we used the default values.
This includes generating}
X-ray light curves in two bands: hard (HX; \rb{1.5}--10~keV) and soft X-rays (SX; 0.3--\rb{1.5}~keV).
\rb{For a detailed discussion of this tool and the default parameter values, please see \cite{Evans09}.}

Note that the XRT has two observing modes: photon counting (PC) and windowed timing (WT).
The vast majority of these observations were made in PC mode.
In order to create uniform X-ray datasets for time-series analysis, we restrict light curve measurement to the single best-used mode for each target, so the small amount of WT data were ignored.

\subsection{Light Curves}
\label{section:lcs}

The uniform reduction we have described was performed on these four datasets in order to allow consistent time-series analysis, both in this paper and more broadly by the community.
These light curves are plotted in \rb{Figure~1}.
These reduced UVOT and XRT data are also compiled in a single table, \rb{Table~2}, for ease of use.
This table is available online through {\it the Astrophysical Journal} in machine readable format, as well as at the Digital Repository at the University of Maryland (DRUM).\footnote{These IDRM data can be downloaded at \url{https://drum.lib.umd.edu/handle/1903/21536}\label{DRUM}.}
The main advance over the three sets of light curves presented previously (\nga, \citealt{Edelson15}; \ngb, \citealt{Edelson17}; \ngc, \citealt{McHardy18}) is the superior rejection of UVOT dropouts. 
Note that all four of these datasets are well-suited for IDRM: all show strong variability in all UVOT bands as well as the hard X-ray band, and most show measurable variability in the soft X-rays as well.

It is visually apparent that for each object the UV/optical light curves are all quite similar, showing relatively slow variations that seem to be adequately sampled at these high IDRM sampling rates. 
By comparison the X-rays show higher amplitude variability on the shortest timescales sampled, and perhaps even with these high sampling rates the variations are undersampled.
Finally it is clear that the X-ray/UV relationship is more complicated than that within the UV/optical.
These relationships will be quantified and discussed in the following sections.

\begin{center}
\begin{deluxetable}{lcccccc}
\label{table2}
\tablenum{2}
\tablecaption{Data}
\tablewidth{0pt}
\tablecolumns{7}
\tablehead{
\colhead{(1)} & \colhead{(2)} & \colhead{(3)} & \colhead{(4)} & 
\colhead{(5)} & \colhead{(6)} & \colhead{(7)} \cr
\colhead{Object} & \colhead{Filter} & \colhead{Cad.} & \colhead{MJD} & 
\colhead{Dur.} & \colhead{Flux} & \colhead{Error} } 
\startdata
Mrk~509 & W1 & 1001 & 57829.8521 & 161.8 & 4.601 & 0.067 \cr
Mrk~509 & U & 1001 & 57829.8535 & 80.8 & 3.058 & 0.056 \cr
Mrk~509 & B & 1001 & 57829.8545 & 80.8 & 1.655 & 0.031 \cr
Mrk~509 & HX & 1001 & 57829.8558 & 990.4 & 0.924 & 0.081 \cr
Mrk~509 & SX & 1001 & 57829.8558 & 990.4 & 1.376 & 0.099 \cr
Mrk~509 & W2 & 1001 & 57829.8569 & 323.8 & 5.930 & 0.073 \cr
Mrk~509 & V & 1001 & 57829.8593 & 80.8 & 1.264 & 0.030 \cr
Mrk~509 & M2 & 1001 & 57829.8612 & 237.3 & 5.110 & 0.076 \cr
Mrk~509 & W1 & 1002 & 57830.9084 & 155.8 & 4.527 & 0.066 \cr
Mrk~509 & U & 1002 & 57830.9098 & 77.8 & 3.194 & 0.059
\enddata
\tablecomments{Column 1: Object name.
Column 2: Filter/band used to measure the data point.
Column 3: Cadence number, where the most significant digit refers to the object and the next three refer to the visit number for that object.
Column 4: Modified Julian Day at the midpoint of the exposure.
Column 5: Duration of the integration in that filter/band, in seconds.
Column 6: Mean flux of the data point.  UVOT fluxes are given in units of  $10^{-14}$ erg cm$^{-2}$ s$^{-1}$ \AA$^{-1}$ and X-ray fluxes  in units of ct s$^{-1}$.
Column 7: Uncertainty on the flux, in the same units as Column~5.
Only a portion of this table is shown here to demonstrate its form and content. 
A machine-readable version of the full table is available online.}
\end{deluxetable}
\end{center}

\section{Time-series Analysis}
\label{section:tsa}

\subsection{Variability Amplitudes}
\label{section:fv}

The fractional variability $F_\mathrm{var}$ \citep{Vaughan03} was used to quantify the variability amplitude in each band.
($F_\mathrm{var} = \sqrt{S^2-\sigma^2_\mathrm{err}}/{\langle X \rangle}$, where $\langle X \rangle$ and $S$ are the mean and total variance of the light curve and $\sigma^2_\mathrm{err}$ is the mean error.)
This is given in Column~4 of \rb{Table~3}.  
It is clear that within the UV/optical, the fractional variability amplitude generally decreases with increasing wavelength, consistent with the longer wavelength emission arising from larger disk annuli, and dilution by a relatively red spectrum including host galaxy starlight, as has been widely reported.
Within the X-rays the situation is not as consistent.
For \ngb, the hard X-rays are strongly variable but the soft X-rays are only weakly variable, while in the other three IDRM AGN the soft X-rays show larger fractional variability than the hard X-rays.
The implications of this behavior will be discussed further in Section~4.

\subsection{Cross-correlation Analyses}
\label{section:ccf}

The focus of this paper is on testing and constraining continuum-emission models  through measurement of interband lags.
We used two methods to measure the interband correlation and lags: the interpolated cross-correlation function (ICCF; \rb{\citealt{Gaskell87}}) and \javelin\ \citep{Zu11}.
We discuss each of these below.

\subsubsection{Interpolated Cross Correlation Function}
\label{section:iccf}

We used the {\tt sour} \rb{code} of the ICCF, \rb{which is based on the specific implementation of the ICCF presented in \cite{Peterson04}}.\footnote{This code is available at \url{https://github.com/svdataman/sour}}
We first normalized the data by subtracting the mean and dividing by the standard deviation.
These were derived ``locally'' --- only the portions of the light curves that are overlapping for a given lag are used to compute these quantities. 
We implemented ``2-way'' interpolation, which means that for each pair of bands we first interpolated in the ``reference'' band and then measured the correlation function, next interpolated in the ``subsidiary'' band and measured the correlation, and subsequently averaged the two to produce the final cross-correlation function (CCF).
The W2 light curve is always the reference and the other seven bands are considered the subsidiary bands in this analysis.
This band was chosen because it has the shortest UV wavelength and thus is closest to the thermal peak of the accretion disk, in spite of the fact that it has higher leakage than the cleaner (but longer wavelength) M2 band.
\rb{The CCF ($r(\tau)$ where $\tau$ is the lag) is then measured and presented to the right of the light curves in Figure~1.}

\subsubsection{Comparison of $r_\mathrm{max}$ in different bands}
\label{section:rmax}

\rb{The most important parameter derived from the CCF is $r_\mathrm{max}$, the maximum value obtained for the correlation coefficient $r(\tau)$.
This is because if the two bands are not intrinsically correlated, then the interband lag (discussed in the next section) has no meaning.
This quantity is given in Column~5 of Table~3.}

\begin{center}
\begin{deluxetable*}{lccccccccc}
\label{table3}
\tablenum{3}
\tablecaption{Variability Amplitude and Interband Correlation Results}
\tablewidth{0pt}
\tablecolumns{9}
\tablehead{
\colhead{(1)} & \colhead{(2)} & \colhead{(3)} & \colhead{(4)} & 
\colhead{(5)} & \colhead{(6)} & \colhead{(7)} & \colhead{(8)} & 
\colhead{(9)} \cr
& & & & \colhead{ICCF} & \colhead{ICCF} & \colhead{\javelin} & \colhead{Diff. in} & 
  \colhead{\javelin/ICCF} \cr
\colhead{Object}  & \colhead{Band} & \colhead{N} & \colhead{$ F_\mathrm{var}$} & 
  \colhead{$r_{\rm max}$} & \colhead{$\tau_{\rm med}$} & \colhead{$\tau_{\rm med}$} & 
	\colhead{median lags} & \colhead{error ratio} \cr
& & & \colhead{(\%)} & &  \colhead{(days)} & \colhead{(days)} & 
	\colhead{(/$\sigma_\mathrm{Tot}$)} & }
\startdata
Mrk 509 & HX & 254 & 21.7 & 0.633 & 4.941 +2.020/$-$1.390 & \cr
Mrk 509 & SX & 254 & 29.3 & 0.768 & 2.332 +0.850/$-$0.878 & \cr
Mrk 509 & W2 & 233 & 23.3 & 1.000 & 0.000 +0.542/$-$0.547 & \cr
Mrk 509 & M2 & 221 & 21.7 & 0.997 & $-$0.047 +0.552/$-$0.544 & $-$0.172 +0.115/$-$0.129 & 0.223 & 0.223 \cr
Mrk 509 & W1 & 227 & 17.5 & 0.996 & $-$0.047 +0.624/$-$0.598 & 0.005 +0.147/$-$0.148 & $-$0.083 & 0.241 \cr
Mrk 509 & U & 245 & 16.6 & 0.982 & 2.626 +0.583/$-$0.586 & 1.959 +0.219/$-$0.233 & 1.064 & 0.387 \cr
Mrk 509 & B & 245 & 13.9 & 0.980 & 1.937 +0.638/$-$0.616 & 1.548 +0.285/$-$0.261 & 0.569 & 0.435 \cr
Mrk 509 & V & 238 & 10.7 & 0.970 & 2.469 +0.804/$-$0.754 & 2.776 +0.402/$-$0.386 & $-$0.352 & 0.506 \cr
\hline
NGC 5548 & HX & 268 & 27.3 & 0.385 & $-$4.550 +1.189/$-$0.720 & & & \cr
NGC 5548 & SX & 268 & 50.6 & 0.438 & $-$2.008 +0.439/$-$0.408 & & & \cr
NGC 5548 & W2 & 260 & 17.5 & 1.000 & $-$0.001 +0.147/$-$0.147 & & & \cr
NGC 5548 & M2 & 249 & 16.6 & 0.993 & $-$0.007 +0.167/$-$0.172 & 0.012 +0.030/$-$0.028 & $-$0.110 & 0.171 \cr
NGC 5548 & W1 & 261 & 13.8 & 0.988 & 0.301 +0.184/$-$0.171 & 0.102 +0.077/$-$0.058 & 1.048 & 0.380 \cr
NGC 5548 & U & 267 & 12.5 & 0.976 & 1.146 +0.174/$-$0.175 & 0.974 +0.094/$-$0.092 & 0.870 & 0.533 \cr
NGC 5548 & B & 271 & 9.2 & 0.965 & 1.108 +0.232/$-$0.237 & 0.980 +0.125/$-$0.141 & 0.475 & 0.567 \cr
NGC 5548 & V & 263 & 6.1 & 0.928 & 1.410 +0.428/$-$0.407 & 1.266 +0.263/$-$0.262 & 0.292 & 0.629 \cr
\hline
NGC 4151 & HX & 314 & 36.4 & 0.677 & $-$3.324 +0.268/$-$0.350 & & & \cr
NGC 4151 & SX & 314 & 10.6 & 0.363 & $-$2.408 +1.461/$-$3.129 & & & \cr
NGC 4151 & W2 & 251 & 6.1 & 1.000 & 0.000 +0.255/$-$0.255 & & & \cr
NGC 4151 & M2 & 250 & 5.8 & 0.973 & 0.055 +0.248/$-$0.239 & 0.045 +0.070/$-$0.053 & 0.040 & 0.253 \cr
NGC 4151 & W1 & 268 & 5.6 & 0.954 & $-$0.011 +0.251/$-$0.264 & 0.064 +0.122/$-$0.113 & $-$0.265 & 0.456 \cr
NGC 4151 & U & 310 & 6.0 & 0.943 & 0.679 +0.239/$-$0.239 & 0.443 +0.162/$-$0.178 & 0.805 & 0.711 \cr
NGC 4151 & B & 311 & 3.0 & 0.895 & 0.877 +0.326/$-$0.352 & 0.475 +0.198/$-$0.205 & 1.019 & 0.594 \cr
NGC 4151 & V & 303 & 2.3 & 0.822 & 0.960 +0.505/$-$0.497 & 0.714 +0.386/$-$0.385 & 0.389 & 0.769 \cr
\hline
NGC 4593 & HX & 191 & 30.1 & 0.690 & $-$0.602 +0.114/$-$0.121 & & & \cr
NGC 4593 & SX & 191 & 34.7 & 0.725 & $-$0.538 +0.101/$-$0.145 & & & \cr
NGC 4593 & W2 & 148 & 12.7 & 1.000 & 0.000 +0.073/$-$0.073 & & & \cr
NGC 4593 & M2 & 149 & 11.3 & 0.971 & 0.048 +0.085/$-$0.086 & 0.009 +0.021/$-$0.021 & 0.443 & 0.246 \cr
NGC 4593 & W1 & 151 & 9.1 & 0.961 & 0.077 +0.110/$-$0.117 & $-$0.010 +0.047/$-$0.040 & 0.716 & 0.383 \cr
NGC 4593 & U & 180 & 7.2 & 0.936 & 0.337 +0.106/$-$0.108 & 0.337 +0.068/$-$0.072 & 0.000 & 0.654 \cr
NGC 4593 & B & 181 & 3.8 & 0.850 & 0.182 +0.172/$-$0.177 & 0.041 +0.113/$-$0.051 & 0.731 & 0.470 \cr
NGC 4593 & V & 176 & 2.2 & 0.701 & 0.351 +0.271/$-$0.298 & 0.182 +0.442/$-$0.138 & 0.416 & 1.019 
\enddata
\tablecomments{Column 1: Object.
Column 2: Band.
Column 3: Number of unique good visits in that band.
Column 4: $ F_\mathrm{var}$, the fractional variability amplitude in that band.
Column 5: ICCF maximum correlation coefficient.
Column 6: Median lag and 68\% confidence interval determined by the ICCF FR/RSS technique.
Here and in \rb{Figures~1 and 3}, a positive value means the comparison band lags behind the reference band (W2).
Note that these are all observed-frame lags, not corrected for time dilation.
Column 7: Median lag and 68\% confidence interval determined by the \javelin\ technique.
Column 8: Difference between the \javelin\ and ICCF median lags, divided by the total 1$\sigma$ uncertainties.
Here we define $\sigma_\mathrm{Tot} = \sqrt{\sigma_\mathrm{ICCF}^2 + \sigma_\mathrm{Javelin}^2}$.
Column 9: Ratio of uncertainties produced by \javelin\ and ICCF FR/RSS (from Columns 6 and 7).
Note that all correlations are measured relative to the \swift\ W2 band, so the W2 lines refers to the autocorrelation; all others are cross-correlations.}
\end{deluxetable*}
\end{center}

\rb{This survey allows quantitative comparison of the level of correlation within the UV/optical with that between the X-rays and UV, because we can use the distribution of $r_\mathrm{max}$ to estimate the sample means and standard deviations for each lag pair.}
The lamp-post model holds that the observed X-ray variability drives that in the UV/optical, at least in its simplest manifestation.
Thus, it predicts strong correlations between the observed X-ray and UV light curves, at least as strong as those observed between the UV and optical.

\rb{Figure~2} plots the measured values of $r_\mathrm{max}$ \rb{(given in Column~5 of Table~3)} for each of these IDRM AGN in the first four panels.
The fifth panel shows the derived mean and standard deviations of $r_\mathrm{max}$ in each band.
It is apparent that X-rays show much weaker correlations with the UV (W2) than is seen between the longer-wavelength UV and optical bands.
\rb{We performed a Kolmogorov-Smirnov (K-S) test to compare the distribution of $r_\mathrm{max}$ for the eight X-ray/UV cases (HX and SX for four targets) and 20 UV/optical ones (five bands [W2/W2 was excluded as that value of $r_\mathrm{max}$ is identically unity] for four targets).
The two-sided K-S test yielded a probability value of $ 7 \times 10^{-5} $, indicating at high confidence that these two samples are not drawn from the same parent population.}
This large difference \rb{in $r_\mathrm{max}$} is not what would be expected from the simple lamp-post reprocessing model, as discussed in Section~4.5.

\subsubsection{\rb{Interband lag measurement and error estimation using FR/RSS}}
\label{section:rmax}

We then used the ``flux randomization/random subset selection'' (FR/RSS) method \citep{Peterson98} to estimate uncertainties on the measured lags.
This is a Monte Carlo technique in which lags are measured from multiple realizations of the CCF.
The FR aspect of this technique perturbs in a given realization each flux point consistent with the quoted uncertainties assuming a Gaussian distribution of errors.
In addition, for a time series with $N$ data points, the RSS randomly draws with replacement $N$ points from the time series to create a new time series.
In that new time series, the data points selected more than once have their error bars decreased by a factor of $n_\mathrm{rep}^{-1/2}$, where  $n_\mathrm{rep}$ is the number of repeated points. 
Typically a fraction of $\left(1-{\frac{1}n}\right)^n\rightarrow 1/e$ of data points are not selected for each RSS realization. 
In this paper, the FR/RSS is applied to both the ``reference'' and subsidiary light curves in each CCF pair.
The CCF ($[r(\tau)]$ where $\tau$ is the lag) is then measured and a lag determined to be the weighted mean of all points with $ r > 0.8 r_\mathrm{max} $, where $ r_\mathrm{max} $ is the maximum value obtained for the correlation coefficient $r$, given in Column~5 of \rb{Table~3}.

For the data presented herein, lags are determined for 25,000 realizations and then used to derive the median centroid lag and 68\% confidence intervals, shown in Column~6 of \rb{Table~3.
This number of trials was chosen so that the uncertainties on the derived median lags and confidence intervals due to Poisson statistics would be negligible compared to that due to the sampling properties and data themselves.
Repeating this test confirms that these quantities only change by very small amounts compared to the widths of the confidence intervals.}

\subsubsection{\javelin}
\label{section:javelin}

We have also employed a second technique, \javelin\ \citep{Zu11}, to estimate the interband lags.  
Rather than linearly interpolating between gaps, \javelin\ models the light curves using a Markov chain Monte Carlo approach.  
The two basic assumptions made by \javelin\ are that the driving light curve is well-modeled by a Damped Random Walk (DRW), and that the other light curves are related to it via a transfer function.  
The standard implementation of \javelin\ assumes a top-hat transfer function (with the top-hat width a free parameter).
Fitting the light curves with \javelin\ begins by modeling the reference light curve with a DRW model.  
The power spectrum of a DRW (see Equation 2 in \citealt{Kelly09}) is equivalent to a \rb{PSD} with a slope of $-$2 at high frequencies ($f > \left[ 2\pi \tau \right]^{-1}$, where $\tau$ is the relaxation time), and flattens off to a constant below this frequency.  
After fitting the reference light curve, other light curves are subsequently fitted assuming the reference light curve model is shifted and blurred by the transfer function.

As with the ICCF analysis we use W2 as the reference band when determining the interband lags. 
Moreover we use the standard top-hat transfer function within \javelin.  
\javelin\ assumes that the higher energies drive the lower energies, and it does not measure the equivalent of an autocorrelation.
For this reason no \javelin\ results are given for the three highest energy bands.
The \javelin\ lags for the five lowest energy bands with W2 are given in Column~7 of \rb{Table~3}.

\subsubsection{Comparison of FR/RSS and \javelin\ uncertainties}
\label{section:comparison}

These ICCF FR/RSS and \javelin\ results are compared in Columns~8 and 9 of \rb{Table~3}.
The median lags are generally quite consistent, all within 1.1$\sigma$ of each other.
However the uncertainties are not consistent; in all but one case the \javelin\ uncertainties are much smaller than the ICCF FR/RSS uncertainties, often by a factor of a few.
In order to explore this further we combine these data with the two others that report both ICCF FR/RSS and \javelin\ results for IDRM AGN, referenced to an ultraviolet band: \cite{Fausnaugh16} on \hst/\swift/ground-based monitoring of \nga\ and \cite{McHardy18} on \swift\ monitoring of \ngc.
These data are ideal for comparison of the two techniques, as they involve consistent application of both to many light curve pairs.
\rb{Figure~4} shows the \javelin\ uncertainties plotted as a function of the ICCF FR/RSS uncertainties on the same light curve pair.
The fitted line indicates that on average the \javelin\ uncertainties are a factor of $\sim$2.5 smaller than the corresponding ICCF FR/RSS uncertainties.

It is unclear why the uncertainties produced by \javelin\ are smaller than those estimated by cross-correlation techniques.
\cite{McHardy18} suggest that this may be due to the actual interband transfer function not being adequately described as a top-hat function, the default for \javelin.
A second possible explanation relates to \javelin's assumption that the PSD slope is equal to or flatter than -2, while the best-sampled observed AGN optical PSDs appear to be steeper that this (e.g. \citealt{Mushotzky11}, \citealt{Edelson14}).
A third is that \javelin\ assumes the errors are Gaussian; the observation of dropouts in the UVOT suggests that this is not the case with these data.

Similarly there is some indication that the ICCF FR/RSS technique could be overestimating the errors.
\cite{Cackett18} applied this technique to the \ngc\ \swift\ and \hst\ data and found that for the well-sampled \swift\ data only, restricting the analysis to just the FR step yielded yielded errors that were a factor of $\sim$2 smaller than that obtained by including both steps.
An even larger discrepancy was apparent when the joint \swift/\hst\ data were analyzed.
Thus it is not clear at this stage to what degree each method is responsible for this discrepancy.

We are planning a comprehensive examination of this issue, which is beyond the scope of the current work.
A more immediate question is which uncertainties to use at this time given that \javelin\ returns uncertainties that are on average only $\sim$40\% the size of those returned by the ICCF FR/RSS technique.
Due to the fact that the older ICCF FR/RSS technique has been more extensively tested, and in order to make more conservative claims about the statistical significance of our lag detections, we restrict the following analyses to results obtained by the ICCF FR/RSS technique.

\section{Discussion}
\label{section:disc}

Now that IDRM has been performed on this small sample, one can look for systematic trends between  variations in different AGN and in different continuum bands.
Section~4.1 describes the characteristics of the variability, Section~4.2 reports the results of fits to test the $ \tau - \lambda $ relation, Section~4.3 uses these results to estimate source parameters of the putative accretion disk, Section~4.4 discusses the large lag excesses seen in {\it U} band and their implications for emission from the BLR, and Section~4.5 discusses the implications of the apparent disconnect between the X-ray and UV light curves.

\subsection{Summary of observed multiband variability}
\label{section:summary}

Visual examination of the light curves shown in \rb{Figure~1} indicates a strong consistency within the UV/optical regime.
For each object, the UV/optical variations all look qualitatively similar, modified by the fact that the {\it V} band variations are more difficult to discern due to the lower intrinsic variability and larger dilution from the (constant) galaxy at longer wavelengths, and that the \swift\ UVOT is relatively less sensitive in {\it V}.
The CCFs show a tendency for the interband lags to increase to longer wavelengths, as will be quantified in the next subsection.
An obvious exception is the {\it U} band, which tends to show longer lags than would be expected from interpolation between {\it B} and W1.
This is likely due to contamination by line and diffuse continuum emission from the (larger) BLR, as discussed in Section~4.4.
Finally, as is apparent in \rb{Figure~1}, there is a general trend for the least luminous/massive targets to show the most rapid variability (\citealt{VandenBerk04}, \citealt{McHardy06}), e.g. the least luminous/massive target in this sample, \ngc, shows the fastest UV/optical variations while the most luminous/massive one, \mrk, shows the slowest variability.
The remaining two targets, \nga\ and \ngb, have similar luminosities and masses, and exhibit intermediate variability.
As discussed in Sections~4.2 and 4.3, this broadly similar UV/optical behavior appears to be consistent with the standard centrally illuminated thin accretion disk model, modified by contamination from the BLR continuum in the {\it U} band, although the disk sizes appear to be somewhat larger than predicted.

Unlike the situation within the UV/optical, no clear trends are apparent within the X-rays or between the X-rays and UV/optical.
Visually for each object, the X-ray variations appear to be more rapid than in the UV/optical bands.
\rb{Figure~2 shows quantitatively} that the peak cross-correlation coefficients, $r_{max}$, are generally much lower between the X-rays and UV than within the UV/optical.
Further, in three cases the hard X-rays \rb{(HX)} are seen to lead \rb{the UV (W2)} by at least 1$\sigma$, while in the other (\mrk), \rb{HX appears} to lag behind W2.
However it is difficult to say if this is real because of the dissimilarity between the X-ray and UV light curves (as evidenced by their low values of $r_{\mathrm max}$).
As discussed in Section~4.5, \rb{the poor UV/X-ray correlations and lack of visual similarity between the UV and X-ray light curves are very} hard to understand in terms of the standard reprocessing picture.

\subsection{Interband Lag Fits}
\label{section:fits}

\rb{Figure~5} plots the interband lag ($\tau$) as a function of continuum wavelength ($\lambda$).
This analysis broadly follows the methodology of \cite{Edelson15} and \cite{Edelson17}.
The standard centrally illuminated thin disk/reprocessing model predicts they should be related by $ \tau \propto \lambda^{4/3} $ \citep{Cackett07}.
This was tested by fitting these data with the function 
$ \tau = \tau_0 [ (\lambda/\lambda_0)^{4/3} - 1 ] $,
where $\lambda_0 = 1928 $~\AA, the central wavelength of the reference W2 band, and $\tau_0$ is the fitted lag between wavelength zero and $\lambda_0$.
The W2 autocorrelation function lag is identically zero, so this point does not participate in the fit but instead the fit is forced to pass through this point.
These data were also fitted with the more general function 
$ \tau =  \tau_0 [ (\lambda/\lambda_0)^{\alpha} - 1 ] $,
where $\alpha$ is the power-law index.
The fit results are shown in \rb{Table~4}.

\begin{center}
\begin{deluxetable*}{lccccccccc}
\label{table4}
\tablenum{4}
\tablecaption{$\tau - \lambda$ Fitting Results}
\tablewidth{0pt}
\tablecolumns{10}
\tablehead{
\colhead{(1)} & \colhead{(2)} & \colhead{(3)} & \colhead{(4)} & \colhead{(5)} & 
 & \colhead{(6)} & \colhead{(7)} & \colhead{(8)} & \colhead{(9)}  \cr
& & \multicolumn{3}{c}{Model 1: $ \tau = \tau_0 [ (\lambda/\lambda_0)^{4/3} - 1 ] $ } 
 & & \multicolumn{4}{c}{Model 2: $ \tau = \tau_0 [ (\lambda/\lambda_0)^\alpha - 1 ] $} \cr
\cline{3-5} \cline{7-10}
Target & dataset & $\tau_0$ (days) & $\chi^2$/dof & p & & $\tau_0$ (days) & 
	$\alpha$ & $\chi^2$/dof & p }
\startdata
Mrk 509	&	Full	&	0.82 $\pm$ 0.17	&	35.90/6	&	$<$0.0001	&	&	0.25 $\pm$ 0.19	&	2.48 $\pm$ 0.68	&	30.77/5	&	$<$0.0001	\cr
Mrk 509	&	UVOT	&	0.85 $\pm$ 0.19	&	0.95/3	&	0.814	&	&	0.44 $\pm$ 0.92	&	1.88 $\pm$ 1.83	&	0.86/2	&	0.6516	\cr
NGC 5548	&	Full	&	0.70 $\pm$ 0.07	&	39.84/6	&	$<$0.0001	&	&	2.83 $\pm$ 0.53	&	0.44 $\pm$ 0.09	&	12.36/5	&	0.0302	\cr
NGC 5548	&	UVOT	&	0.51 $\pm$ 0.09	&	0.85/3	&	0.8376	&	&	0.68 $\pm$ 1.14	&	1.12 $\pm$ 1.19	&	0.82/2	&	0.6645	\cr
NGC 4151	&	Full	&	0.66 $\pm$ 0.10	&	73.77/6	&	$<$0.0001	&	&	4.64 $\pm$ 1.32	&	0.19 $\pm$ 0.09	&	1.71/5	&	0.8873	\cr
NGC 4151	&	UVOT	&	0.35 $\pm$ 0.12	&	0.81/3	&	0.8483	&	&	0.21 $\pm$ 0.54	&	1.76 $\pm$ 2.20	&	0.77/2	&	0.6815	\cr
NGC 4593	&	Full	&	0.27 $\pm$ 0.04	&	18.75/6	&	0.0046	&	&	0.61 $\pm$ 0.11	&	0.52 $\pm$ 0.17	&	2.12/5	&	0.8322	\cr
NGC 4593	&	UVOT	&	0.11 $\pm$ 0.06	&	0.16/3	&	0.9834	&	&	-2.80 $\pm$ 20.29	&	-0.10 $\pm$ 0.74	&	0.12/2	&	0.9425	
\enddata
\tablecomments{Column 1: Target name.
Column 2: Indication of whether the fit included all UVOT/XRT data (full) or just the four UVOT bands excluding U (UVOT).
Columns 3-5: Derived fit parameter $\tau_0$, $\chi^2$/degrees of freedom, and {\it p}-value for Model 1.
Columns 6-9: Derived fit parameters $\tau_0$ and $\alpha$, $\chi^2$/degrees of freedom, and {\it p}-value for Model 2.
Results for each target are given in pairs of rows, the first covering all bands and the second just the  UVOT bands.
}
\end{deluxetable*}
\end{center}

In general, these full data are not well fitted by these functions for two reasons.
First, the X-ray data are incongruent, often falling above or below the fitted lines by many $\sigma$.
Second, excess lags are seen in {\it U} band for all targets.
The reasons for these two complications are discussed in Sections~4.5 and 4.4, respectively.
Thus, \rb{Figure~5} also presents a second set of fits performed after eliminating the X-ray and {\it U} band data so as to just focus on the lag spectrum due to the putative accretion disk.
These data are well fitted, indicating that the standard centrally illuminated thin accretion disk model can account for the overall shape of the $ \tau - \lambda $ relation.
However, in the most extreme case the fit has only 2 degrees of freedom (dof), so this cannot in itself be considered a particularly strong test of the model.
Note also that in two of the four cases (\mrk\ and \nga) of UVOT-only fitting, the Model 2 fits (in which the power-law index is allowed to vary) are significantly better than those for Model 1, in which the power-law index is fixed at 4/3.  
This too should not be considered too constraining because the derived values of $\alpha$ show no consistent trend, with some larger and some smaller than the predicted 4/3.
Finally, we note that the normalizations measured here and with other \swift-only datasets (\nga: \citealt{Edelson15}, \ngc: \citealt{McHardy18}) are slightly ($\sim 2 \sigma$) smaller than those that include both \swift\ and longer-wavelength data (\nga: \citealt{Fausnaugh16}, \ngc: \citealt{Cackett18}).

\subsection{Accretion disk properties}
\label{section:disk}

In this subsection we compare the sizes of the accretion disks derived from RM and the reprocessing model with theoretical predictions for a standard centrally illuminated thin accretion disk.
The arguments in this paper parallel those given in \cite{Fausnaugh16} and \cite{Edelson17}.
The reprocessing model holds that the UV/optical variations are driven by the relatively small, variable, centrally located X-ray emitting corona.
Likewise the $\tau - \lambda$ fits described above were derived using the assumptions of the standard thin disk accretion disk model.
If so, then the parameter $\tau_0$ derived from the UVOT-only $\tau \propto \lambda^{4/3}$ fits gives an estimate of the light-travel time from the center of the system to the region that emits the 1928~\AA\ (W2 band) light.
This is true even if, as appears to be the case, the observed (relatively low energy) X-rays are not the actual driving light curves.

Under these assumptions, Equation 2 of \cite{Edelson17} gives the light-crossing radius $r$ of an annulus emitting at a characteristic wavelength $\lambda$:
\begin{equation}\label{eq:}
r = 0.09 \left(X\frac{\lambda}{1928{\rm \AA} } \right)^{4/3}\nonumber M_8^{2/3} \left(\frac{\dot m_{\rm Edd}}{0.10} \right)^{1/3}
\textrm{lt-dy}
\end{equation}
where $X$ is a multiplicative scaling factor of order unity that accounts for systematic issues in converting the annulus temperature $T$ to wavelength $\lambda$ at a characteristic radius $R$, $M_8$ is the black hole mass in units of $10^8 \rm M_\odot$ and $\dot{m}_\mathrm{Edd}$ is the Eddington ratio $ L_\mathrm{bol} / L_\mathrm{Edd} $. 
Under the assumption that at an annulus of radius $R$ the observed wavelength corresponds to the temperature given by Wien's Law, then $ X = 4.97 $.
If instead the more realistic flux-weighted radius is used, then $ X = 2.49 $.
(The flux-weighted estimate assumes that the temperature profile of the disk is described by $ T \propto R^{-3/4} $ \citep{Shakura73}.)
In both the Wien and flux-weighted cases, the disk is assumed to have a fixed aspect ratio and to be heated internally by viscous dissipation and externally by the coronal X-ray source extending above the disk.

The result of application of this formula to these data is shown in \rb{Table~5}.
Column~4 gives the ratio of the observed to theoretical centrally illuminated thin accretion disk sizes for the flux-weighted case and Column~6 gives the same ratio for the Wien case.
There is a large spread within each group, which indicates that this ratio is not a terribly consistent determined quantity across this sample.
The median of all of these ratios is a factor of 2.05.
Among the possible causes of this are the large systematic uncertainties in determining $M_{\mathrm{BH}}$ and $\dot m_{\rm Edd}$, both of which could be off by at least a factor of $\sim$3.
Likewise, many of the underlying assumptions, such as the fixed aspect ratio of the disk, are not well established by observation.
Finally, the fit parameter $\tau_0$ is also not well determined with \swift\ data alone; the addition of optical data will provide much better constraints (e.g., \citealt{Fausnaugh16}, \citealt{Cackett18}).
Due to all of these large systematic uncertainties, the present UV/optical interband lags are deemed to be consistent with the predictions of the standard \cite{Shakura73} thin accretion disk model.

\begin{center}
\begin{deluxetable}{lccccc}
\label{table5}
\tablenum{5}
\tablecaption{Accretion disk parameters}
\tablewidth{0pt}
\tablecolumns{6}
\tablehead{
\colhead{(1)} & \colhead{(2)} & \colhead{(3)} & \colhead{(4)} & \colhead{(5)} & 
 \colhead{(6)} \cr
&	$\tau_0$ &	$r_1$ &	Ratio &	$r_2$	&	Ratio  \cr
Target	&	(lt-day)	& (lt-day) & 1 & (lt-day) &	 2 }
\startdata             
Mrk 509	  &	0.85	&	0.260	&	3.28	&	0.65	&	1.31	\cr
NGC 5548	&	0.51	&	0.157	&	3.25	&	0.36	&	1.30	\cr
NGC 4151	&	0.35	&	0.072	&	4.86	&	0.18	&	1.94	\cr
NGC 4593	&	0.11	&	0.051	&	2.14	&	0.13	&	0.85	
\enddata
\tablecomments{Column 1: Target name.
Column 2: Observed value of $\tau_0$ taken from \rb{Table~4}, column 3, even rows, converted to the light-crossing size by assuming $ r = c \tau_0 $.
This is equivalent to the radius of the W2-emitting disk annulus, in lt-days.
Column 3: Theoretical estimate of the flux-weighted radius of the W2-emitting disk annulus in lt-days, assuming $ X = 2.49 $ in Equation 1.
Column 4: Ratio of the observed/theoretical sizes for the flux-weighted case.
Column 5: Theoretical estimate of the Wien radius of the W2-emitting disk annulus in lt-days, assuming $ X =4.97 $ in Equation 1.
Column 6: Ratio of the observed/theoretical sizes for the Wien case.}
\end{deluxetable}
\end{center}

\subsection{Diffuse continuum emission from the BLR}
\label{section:uband}

The bottom panels in \rb{Figure~5} present compelling evidence of ``excess'' lags in {\it U} band, relative to both the $ \tau \propto \lambda^{4/3} $ accretion disk fit and to the surrounding W1 and {\it B} band lags.
This phenomenon was discovered by \cite{Korista01} on the basis of the 1989 {\it IUE} and 1993 \hst\ spectroscopic campaigns.
This result was neglected for the better part of the next decade, in part because of a dearth of campaigns that included both {\it U} and surrounding bands.
That changed when the first IDRM experiment, also on \nga, found an excess lag in {\it U} band that was too large to be ignored (\citealt{Edelson15}, \citealt{Fausnaugh16}). 
Similarly large {\it U}-band excess lags were seen in \swift\ IDRM monitoring of \ngb\ \citep{Edelson17} and \ngc\ \citep{McHardy18}.
The \ngc\ campaign also included \hst\ data, allowing measurement of a lag spectrum with much higher spectral resolution (but lower temporal resolution), which finds a strong excess and discontinuity in the Balmer jump region \citep{Cackett18}.

\cite{Korista01} determined that continuum emission from the BLR contributes significantly to the measured fluxes in the UV-optical continuum windows. 
That study, and more recently \cite{Lawther18}, found that for the range of physical conditions necessary for efficient emission line formation, a significant diffuse continuum component from that same BLR gas is largely unavoidable. 

\rb{Table~6 and Figure~5} make the magnitude of this effect clear.
\rb{The {\it U} band lag shows an excess of a factor of $\sim$2.2 (on average) above those predicted by the model and those derived by interpolation between the observed W2 and {\it B} band lags.
This demonstrates quantitatively that the BLR continuum component must contribute significantly to the observed lags, and observation of this strong excess in all four of the targets surveyed suggests it is a common occurrence in AGN.}
However, the disk (or some similarly compact component, relative to the BLR) must also contribute as otherwise the observed \rb{continuum interband lags throughout the UV/optical} would be comparable to those measured in the broad emission lines.
\rb{While this analysis shows that both the BLR continuum component and the disk must contribute to the UV/optical interband lag spectra of these AGN, as discussed in \cite{Lawther18}, detailed BLR modeling is required to determine the precise contributions of each.
Such modeling is beyond the scope of this paper.}

\rb{\begin{center}
\begin{deluxetable*}{lccccc}
\label{table6}
\tablenum{6}
\tablecaption{{\it U} band lag parameters}
\tablewidth{0pt}
\tablecolumns{6}
\tablehead{
\colhead{(1)} & \colhead{(2)} & \colhead{(3)} & \colhead{(4)} & \colhead{(5)} & 
 \colhead{(6)} \cr
 & {\it U} band & Model 3465 & Model & Int. 3465 & Int. \cr
Target & lag (d) & lag (d) & Ratio & lag (d) & Ratio  }
\startdata
Mrk 509  & 2.63 $\pm$ 0.59 & 1.00 & 2.62 & 0.91 & 2.88 \cr
NGC 5548 & 1.15 $\pm$ 0.17 & 0.61 & 1.88 & 0.69 & 1.66 \cr
NGC 4151 & 0.68 $\pm$ 0.24 & 0.42 & 1.62 & 0.42 & 1.63 \cr
NGC 4593 & 0.34 $\pm$ 0.11 & 0.13 & 2.61 & 0.13 & 2.64
\enddata
\tablecomments{Column 1: Target name.
Column 2: Observed ICCF {\it U} band lag and FR/RSS uncertainty, taken from \rb{Table~3}, in days.
Column 3: Expected lag from Model~1, UVOT-only data (excluding the {\it U} band), evaluated at 3465~\AA, the center of the {\it U} band in days.
Column 4: The ratio of the \rb{observed {\it U} band lag to the expected Model~1 lag (Column 2 divided by Column 3).}
Column 5: Expected lag computed by linear interpolation between the observed W1 and {\it B} band lags evaluated the center of {\it U} band in days.
Column 6: The ratio of the \rb{excess {\it U} band lag to the expected Model~2 lag (Column 2 divided by Column 5).
Note the observed {\it U} band lags are on average twice those expected from the models.}}
\end{deluxetable*}
\end{center}}

Determining the exact mix of the disk and BLR components will require the development of new techniques.
Two types of advances would be helpful:
first, as the original locally optimally emitting cloud (LOC) \cite{Korista01} model of the BLR gas was specific to \nga, more general, robust models, including assumptions that are different from the LOC assumptions, must be made of the BLR gas response over the wide range of conditions seen in AGN.
Because all four of these campaigns also include broadband and spectroscopic monitoring from Las Cumbres Observatory (LCO) and other ground-based observatories, this must include determining the exact structure of contributions across the entire observed UV/optical/IR bandpass.
Second, detailed simulations must be done of the CCF that would emerge from mixing signals at both long (from the BLR) and short timescales (e.g. from the disk).
That is beyond the scope of this paper but should be undertaken urgently, as resolution of this issue is required for fully understanding these and future IDRM data.
Also, future IDRM campaigns could be designed to focus on this by, for example, including \hst\ to provide much higher spectral resolution (e.g. \citealt{Cackett18}).

\subsection{How do the X-Rays fit in?}
\label{section:xrays}

The standard reprocessing model has two main components: 
(1) a geometrically thin, optically thick accretion disk that emits in the UV/optical \citep{Shakura73} and (2) a central X-ray emitting corona \citep{Haardt91} that illuminates and heats the disk.
Likewise, this experiment probes two distinct $2-3$ octave wide regimes separated by about 1.5 orders of magnitude in wavelength: the UV/optical (the UVOT filters cover $\sim 1600-5850$~\AA\ FWHM; \citealt{Poole08}) and the X-rays (the XRT covers $0.3-10$ keV, or $\sim 1.2-40$~\AA).
These two structures are thought to dominate different bands: the disk (as well as the BLR) in the UV/optical and the central corona in the X-rays.
Thus, testing this full picture requires linking the variability of both of these putative emission components, which means bridging this large gap in wavelength.

As discussed above, the variability within the UV/optical is well-understood in terms of the standard \cite{Shakura73} thin, centrally illuminated disk model modified by emission line and diffuse continuum emission from the BLR \citep{Korista01}, although the exact mixing of these two components cannot be measured with these data alone.
However, no such clear pattern emerges within the X-rays, or between the X-rays and the UV/optical.
For three of the IDRM AGN, the HX band shows a significant lead relative to W2, while in the fourth (\mrk) it actually shows a significant lag behind W2.
As the X-ray/UV correlations are generally much weaker (with peak correlation coefficients $r_\mathrm{max} < 0.75 $ in all cases) than those within the UV/optical  ($r_\mathrm{max} > 0.8 $ in all but one case), it is unclear to what extent this is an intrinsic property of AGN variability and to what extent it is an artifact of the CCF analysis.
Likewise, the variability amplitude, as measured by $F_\mathrm{var}$, is much stronger in SX than HX in one source (\nga), much stronger in HX than SX in another (\ngb) and similar in the two X-ray bands in the other two.

Taken as a whole, these results strongly challenge our relatively simple picture of the origin of the X-ray variability.
The central corona reprocessing model (\citealt{Frank02}, \citealt{Cackett07}) would predict that the $ \tau \propto \lambda^{4/3} $ relation seen in the UV/optical (with the exception of the {\it U} band, which is dominated by emission from the BLR) should extrapolate smoothly back to the X-rays, but \rb{Figure~5} clearly demonstrates that this is not the case.
Further this model would predict X-ray/UV correlations that are at least as strong as those between the UV and optical, but \rb{Figure~2} demonstrates \rb{that the opposite is true}.
This disconnect between the observed X-rays and UV is very difficult to reconcile with the reprocessing picture, forcing  us to consider alternate explanations for observed interband variability.

One interesting model is that of \cite{Dexter11}, in which the X-rays are produced in a large number of independently variable regions across the surface of the disk.
However, this model does not currently make clear predictions for the interband lags, so it cannot be tested with these data.

A second model is that of \cite{Gardner17}, which was developed to explain the larger than expected X-ray/UV lags seen in the first IDRM campaign, on \nga.
In this picture the variable central X-ray corona illuminates a geometrically and optically thick ring that extends above/below the disk.
This ring emits in the unobservable extreme ultraviolet (EUV), where it illuminates and heats the disk.
This provides an additional reprocessing step that further smooths and delays the variable signal produced in the directly observable X-ray corona.
However this provides no simple explanation for the \mrk\ result, where the ultraviolet appears to lead the X-rays for at least part of the observations.

More generally, it could be that the 0.3-10 keV X-ray continuum \rb{observed by the \swift\ XRT} is not the same as the driving band that illuminates the disk.
This could be because the driving band is at lower energies (e.g. the EUV, as in the \citealt{Gardner17} picture) or at higher energies, above \rb{the XRT bandpass}.
Or the driving band could be partially obscured from our line of sight, so that the disk sees a different driver than we observe.

A third model \citep{Uttley03} was developed to explain the observation in \nga\ of a stronger optical/X-ray correlation on very long time scales (many years) than in these relatively short campaigns.
This starts with a ``standard'' central corona and adds additional variability on the viscous drift timescale due to infalling matter in the accretion flow.
This ``fueling'' term would dominate the long timescale variability in all bands, leading to the observed strong X-ray/optical correlation observed in \nga\ and other AGN on timescales of many years \citep{Uttley03, Arevalo08, Breedt09, Arevalo09}.
A key prediction of this model is that the red-lags-blue relation reverses to blue-lags-red on long timescales as inward propagation of mass-accretion fluctuations start to dominate.  
However, such a test cannot be performed with the relatively short campaigns reported herein.
It may be testable once LSST produces long ($\sim$5 year) multiband light curves for thousands of AGN.

\section{Conclusions}
\label{section:concl}

This \swift\ survey of the temporal relationships between variations at X-ray, UV, and optical wavelengths in four AGN has clarified our picture of AGN central engines while at the same time raising new questions.
The \rb{first} observational result is that all four AGN show variations that are strongly correlated throughout the UV/optical (\rb{Figure~2}), and all show the same general structure of interband lags increasing from UV to the optical wavelengths (Figure~3a).
After excluding the {\it U} band, \rb{Figure~5} shows that all are well fitted by the $ \tau \propto \lambda^{4/3} $ relationship predicted by the standard thin accretion disk model \citep{Shakura73}, as modified for illumination by a central driver that does not appreciably change the temperature structure of the disk, e.g., by \cite{Cackett07} .
While these lags are a factor of $\sim$2 larger than predicted, the uncertainties on the predicted lags are quite large, so it is not yet clear if this is a problem for the standard thin disk picture.

A second important finding is that the UV/optical interband lag structure is strongly affected by diffuse continuum emission from the BLR, even though these bands do not contain the strongest BLR emission lines.
This is apparent in \rb{Table~6}: the observed lag in the {\it U} band, which contains the 3646~\AA\ Balmer jump, is \rb{on average a factor of $\sim$2.2 above} that expected \rb{both} from interpolating between the surrounding bands and \rb{from the} disk model \rb{fits}.
Theoretical modeling of this ``excess {\it U} band lag'' in one target, \nga, indicates that it is merely the most obvious tracer of lower-level diffuse continuum emission from the BLR that should extend across the UV/optical region observed by \swift\ (\citealt{Korista01}, \citealt{Lawther18}).
Based on the 6-filter UVOT monitoring alone, it is not currently possible to determine the precise mix of disk and BLR emission contributing to the observed lag structure.
Progress in this area will likely require a combination of advances in theory, analytical methods, and experimental design, some of which are discussed below.

A third \rb{key} observational result, which was not generally expected prior to these IDRM campaigns, is that the X-ray variability does not show the strong, consistent link to the UV/optical that is predicted by the reprocessing model.
\rb{Figure~2} shows that the X-ray/UV correlations are much weaker than those within the UV/optical, and Figure~3a indicates a diversity of X-ray/optical lags, with the X-rays leading the UV in three cases and lagging in the fourth.
This poses a severe problem for the reprocessing model, for which no simple solution is currently apparent.

\swift\ is the workhorse without which IDRM cannot (currently) be successfully performed on AGN, as it is the only fully operational observatory that provides coverage throughout the X-ray, UV and optical regimes necessary for this experiment.
However, it is also very helpful to expand coverage to longer wavelengths and higher spectral resolution than that afforded by the UVOT alone.
We note that simultaneous ground-based optical monitoring has been performed on all four of these targets, and the first dataset (on \nga) has already been published \citep{Fausnaugh16}.
Reduction and analysis of optical photometry and spectroscopy on the other three is ongoing and will be published shortly.
This should provide further constraints on the accretion disk fits (especially the overall size) and possibly on the effects of line and diffuse emission from the BLR.
Note as well that \hst\ can also play a crucial role, providing much higher spectral resolution that can allow direct detection and modeling of the \rb{BLR continuum} component, and extending the UV coverage to wavelengths as short as $\sim$1150~\AA.
That has been used to resolve the excess lag around the Balmer jump in one of these targets (\ngc, \citealt{Cackett18}), and it is expected to play a prominent role in future campaigns.

These unprecedented data also reveal the need for improvements in our time-series analysis tools.
The uncertainties on lags output by \javelin\ \citep{Zu11} are on average a factor of $\sim$2.5 smaller than the same quantity output by the ICCF FR/RSS technique \citep{Peterson98}.
The precise cause of this discrepancy has not been determined, but given the fact that the older ICCF FR/RSS technique is more conservative in its assumptions and results, it and not \javelin\ was utilized to derive the results reported above.
More generally, there is no interband lag tool currently in use that allows separation of two distinct interband lag signals, as seems to be the case in the UV/optical, where short lags from the accretion disk and longer lags from the BLR appear to be present.
Ultimately these techniques will need to be developed and directly compared so as to determine which are more reliable and suitable for studying AGN interband lags, but that is beyond the the scope of the current paper.

These results also highlight the need for theoretical progress, especially in understanding the X-ray emission component(s) and relation to the disk and BLR that emit at lower energies.
These observations present severe problems for the ``lamp-post'' reprocessing model.
Can it be modified or must it be discarded?
If it is the latter then what will take its place?
The current set of reduced data are available through {\it The Astrophysical Journal}.
In order to facilitate ongoing modeling and theoretical progress by all interested astronomers, we have also compiled these data at the DRUM archive.\textsuperscript{\ref{DRUM}}
This archive will be updated as improved reduction output (e.g., more comprehensive UVOT dropout filtering; see the Appendix) and current/future IDRM campaign data become available.

\acknowledgments 
The authors note the crucial role played in this research by Neil Gehrels, the late director of \swift: without his decision to allow full 6-filter UVOT monitoring for the duration of these campaigns, these results would not have been possible.
We also appreciate Ian McHardy's leadership of the third IDRM campaign, on \ngc, and Chris Kochanek's input on \javelin.
R.E. and J.M.G. gratefully acknowledge support from NASA under the ADAP award 80NSSC17K0126.
Research by A.J.B. is supported by NSF grant AST-1412693.
K.H. acknowledges support from STFC grant ST/R000824/1.
A. B. and K.P. acknowledge support from the UK Space Agency.
M.C.B. gratefully acknowledges support from the National Science Foundation through CAREER grant AST-1253702.
C.D. acknowledges the Science and Technology Facilities Council (STFC)
through grant ST/P000541/1 for support.
M.V. gratefully acknowledges support from the Independent Research Fund Denmark via grant number DFF 4002-00275.
SRON is supported financially by NWO, the Netherlands Organization for Scientific Research.

\software{{\tt HEAsoft} (v6.22.1; Arnaud 1996), {\tt FTOOLS} (Blackburn 1995), {\tt sour} (Edelson et al. 2017), \javelin\ (Zu et al. 2011)}

\appendix
\label{section:appendix}

\rb{ 
As first noted by \cite{Edelson15}, \swift\ UVOT light curves exhibit occasional ``dropouts,'' anomalously low points most frequently seen in the UV.
Our earlier work indicated  this was due to localized low sensitivity regions (see also \citealt{Breeveld16}).
We identify clusters of dropouts in the detector plane and use these to define detector masks, following the procedure laid out in \cite{Edelson15} and \cite{Edelson17}, except that we now combine data from four AGN, and handle the UV and optical data separately.
Previously, it was noted that dropouts were found less frequently in the {\it U} band and rarely in {\it B} and {\it V}, so UV data were used to define detector masks that were then applied to data in the UV and {\it U} filters.
The present data improve the detector plane coverage, which for the first time makes it possible to identify clusters amongst dropouts in the {\it U}, {\it B} and {\it V} filters, which are found to be less widely distributed than the UV-identified clusters (\rb{Figure~A.1}).
We therefore define two detector masks, one based upon UV dropouts and applied to the three UV filters, the other based upon optical dropouts and applied to the three visible filters.
Table~A.1 summarizes the number of dropouts found in each filter and the result of applying the detector masks to the IDRM data from all four AGN.
Note that column 6 of this table shows that the measurements screened out by the detector masks that do not satisfy our formal definition of dropouts also have systematically low flux values, indicating that these are also affected by the low sensitivity regions.
The mask definitions are presented in Tables~A.2 and A.3.}

\setcounter{table}{0}
\renewcommand{\thetable}{A\arabic{table}}

\begin{deluxetable}{lccccc}
\label{tableA1}
\tablenum{1}
\tablecaption{UVOT dropout data
\label{tab:dropouts}}
\tablecolumns{6}
\tablehead{
\colhead{(1)} & \colhead{(2)} & \colhead{(3)} & 
 \colhead{(4)} & \colhead{(5)} & \colhead{(6)} \cr
\colhead{Filter} & \colhead{Dropout} & \colhead{Dropout} & 
 \colhead{Dropouts}    & \colhead{Non-drop} & \colhead{Non-drop} \cr
\colhead{}  & \colhead{tested} & \colhead{tally} &
 \colhead{in mask} & \colhead{in mask} & \colhead{avg dev}}
\startdata
W2	&	1023	&	129	&	118	&	 16	&	$-$1.86	\cr
M2	&	1186	&	127	&	122	&	 46	&	$-$1.96	\cr
W1	&	1042	&	122	&	 99	&	 38	&	$-$1.29	\cr
U	&	 936	&	 39	&	 24	&	 17	&	$-$0.87	\cr
B	&	1038	&	 16	&	  9	&	 24	&	$-$1.10	\cr
V	&	1005	&	 17	&	 10	&	 20	&	$-$0.60
\enddata
\tablecomments{Column 1: UVOT filter.
Column 2: Number of measurements in intensive monitoring light curves
to which dropout testing is applied.
Column 3: Number of dropouts identified in these light curves.
Column 4: Size of subset of dropouts that fall within detector mask.
Column 5: Number of non-dropout points within mask.
Column 6: Mean deviation of non-dropout points that fall within mask, in units of $\sigma$.}
\end{deluxetable}

\begin{deluxetable}{lcccc}
\label{tableA2}
\tablenum{2}
\tablecaption{UV mask boxes
\label{tab:UV_boxes}}
\tablecolumns{5}
\tablehead{
\colhead{(1)} & \colhead{(2)} & \colhead{(3)} & \colhead{(4)} & \colhead{(5)} \cr
\colhead{Box}  & \colhead{$X_1$} & \colhead{$X_2$} & \colhead{$Y_1$} & \colhead{$Y_2$}}
\startdata
1	&	317	&	326	&	651	&	671	\cr
2	&	331	&	344	&	622	&	627	\cr
3	&	337	&	347	&	662	&	686	\cr
4	&	346	&	350	&	613	&	633	\cr
... \cr
64	&	729	&	736	&	532	&	533
\enddata
\tablecomments{Column 1: Box number.
Columns 2-5: $X$ and $Y$ coordinates of box.
The coordinates in \rb{Tables~A.2 and A.3} are the $X$ and $Y$ ranges spanned by rectangular boxes drawn in the reference frame of raw UVOT images with the default 2$\times$2 binning (with pixels numbered from 0 to 1023).
A machine-readable version of the full table is available online.}
\end{deluxetable}

\begin{deluxetable}{lcccc}
\label{tableA3}
\tablenum{3}
\tablecaption{Optical mask boxes
\label{tab:opt_boxes}}
\tablecolumns{5}
\tablehead{
\colhead{(1)} & \colhead{(2)} & \colhead{(3)} & \colhead{(4)} & \colhead{(5)} \cr
\colhead{Box}  & \colhead{$X_1$} & \colhead{$X_2$} & \colhead{$Y_1$} & \colhead{$Y_2$}}
\startdata
1	&	359	&	377	&	636	&	651	\cr
2	&	416	&	440	&	551	&	558	\cr
3	&	431	&	435	&	650	&	657	\cr
4	&	450	&	463	&	439	&	447	\cr
... \cr
11	&	561	&	577	&	575	&	598
\enddata
\tablecomments{Column 1: Box number.
Columns 2-5: $X$ and $Y$ coordinates of box.
A machine-readable version of the full table is available online.}
\end{deluxetable}


\begin{figure}
\figurenum{1a}
\begin{center}
 \includegraphics[width=7in]{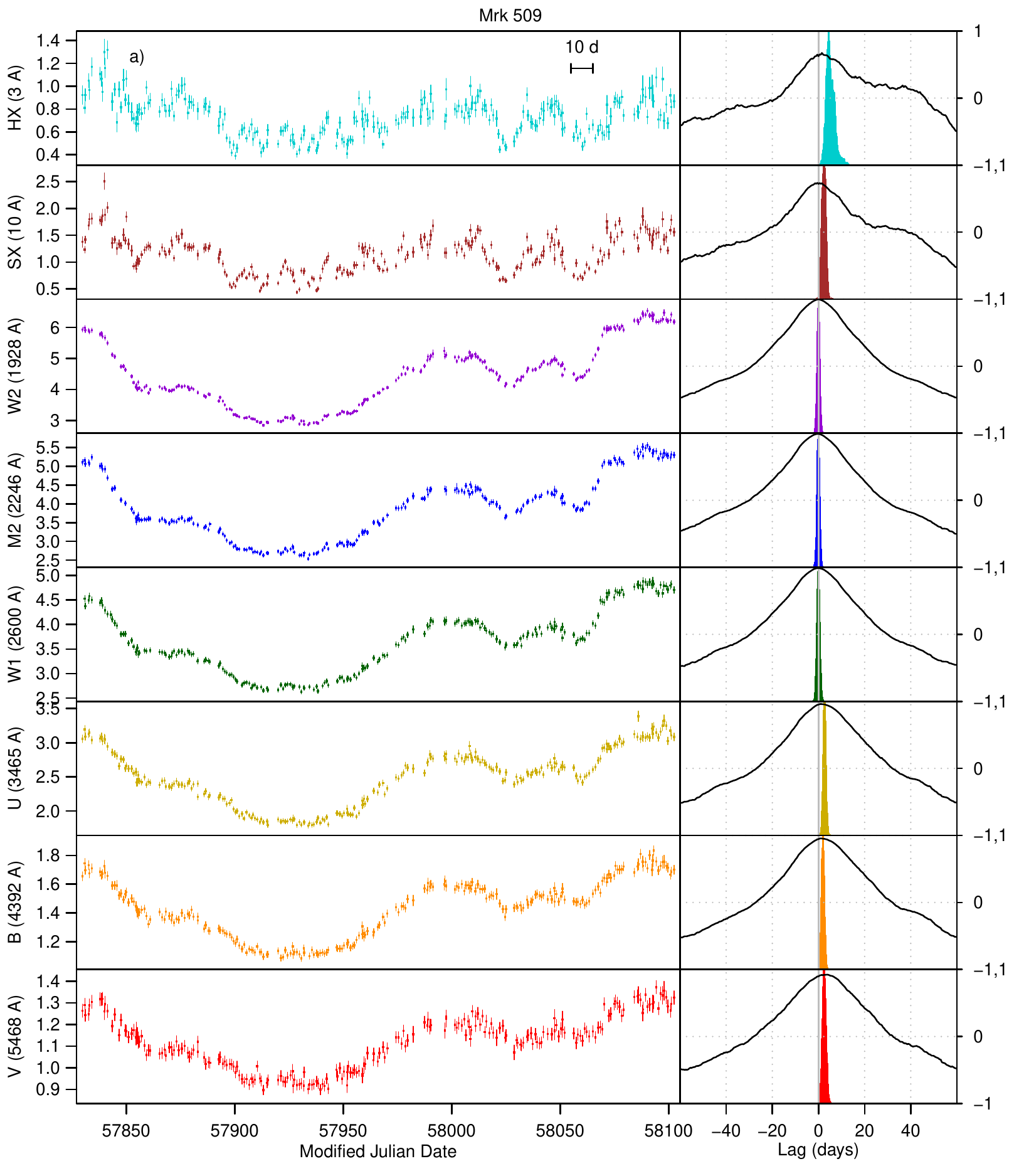} 
\caption{Left: \swift\ IDRM light curve for \mrk.
Data are ordered by wavelength, with the top two panels from XRT (HX covering \rb{1.5}--10~keV and SX 0.3--\rb{1.5}~keV), and the bottom six from UVOT.
The plotted UVOT points are restricted to those that passed the filtering discussed in Section~2.2.
The X-ray data are in units of ct s$^{-1}$ and the UVOT data are in units of $10^{-14}$ erg cm$^{-2}$ s$^{-1}$ \AA$^{-1}$.
Right: ICCFs (in black; scale on the right) and FR/RSS centroid distributions (in the same color as the light curves) for each band relative to the W2 band.
Here and in the remainder of \rb{Figures~1} and 3, a positive value means the comparison band lags behind W2.
The {\it x}-axes cover the full range of lags over which each ICCF is measured and the {\it y}-axes of the ICCF plots cover the full range of $r$ covering $\pm$1.
The vertical gray lines give lags in round numbers of days (solid at zero).
The horizontal dashed gray line shows $ r = 0 $.
The horizontal error bar in the upper right of the top (HX) panel shows 10 days for scale.
\mrk\ was the fourth AGN subjected to \swift\ IDRM in a campaign that covered 2017 March -- December.
The \swift\ data are presented in this paper for the first time.
This AGN has the highest mass ($\sim 1.1 \times 10^8 M_\odot $) of any IDRM target observed to date.}
\label{fig:fig1a}
\end{center}
\end{figure}

\begin{figure}
\figurenum{1b}
\begin{center}
 \includegraphics[width=7in]{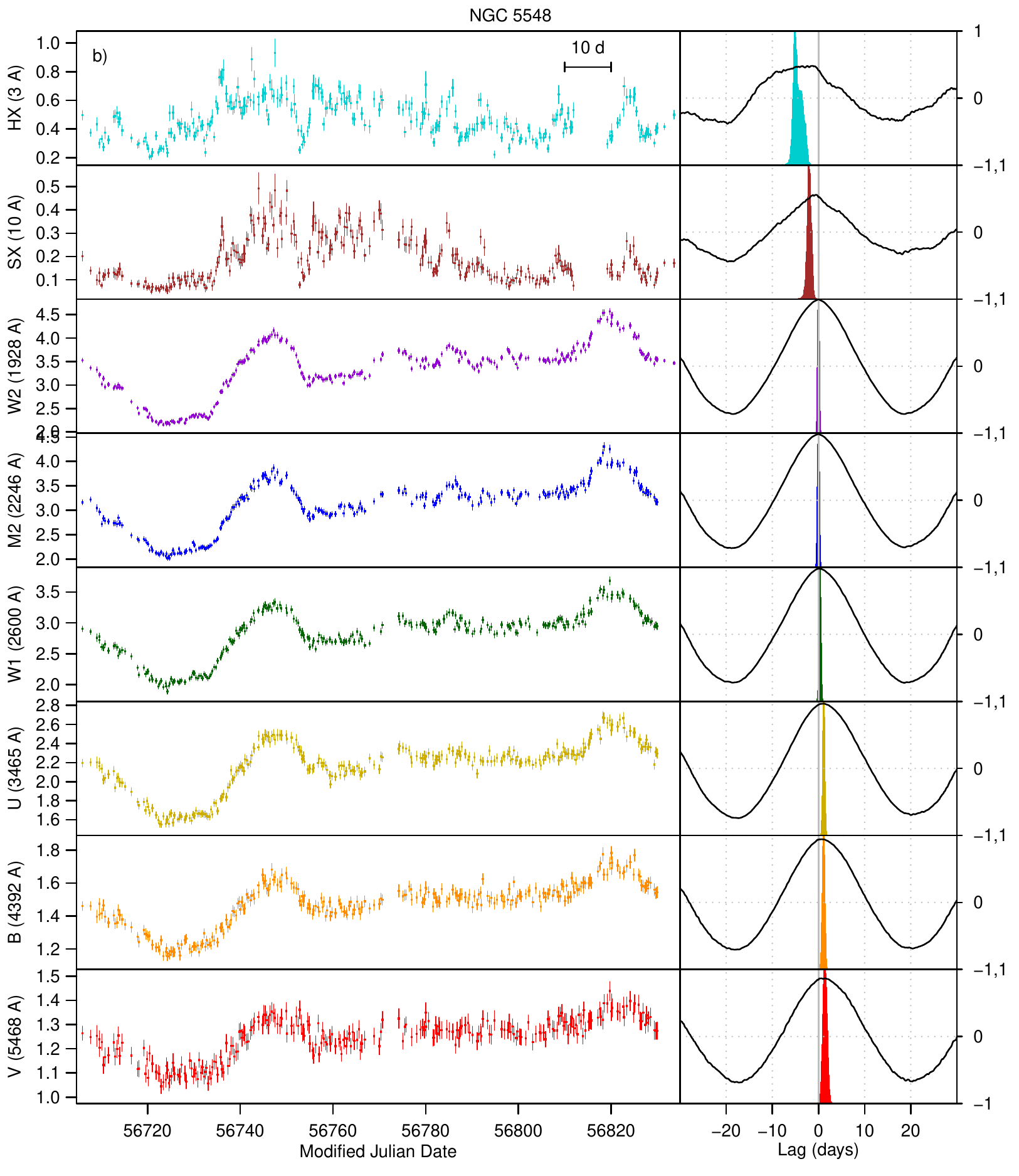} 
\caption{Same as \rb{Figure~1a} except for \nga.
This was the first AGN to be subjected to \swift\ IDRM.
The initial \swift\ analysis, presented in \cite{Edelson15}, showed clear interband lags and established the viability of the IDRM technique.}
\label{fig:fig1b}
\end{center}
\end{figure}

\begin{figure}
\figurenum{1c}
\begin{center}
 \includegraphics[width=7in]{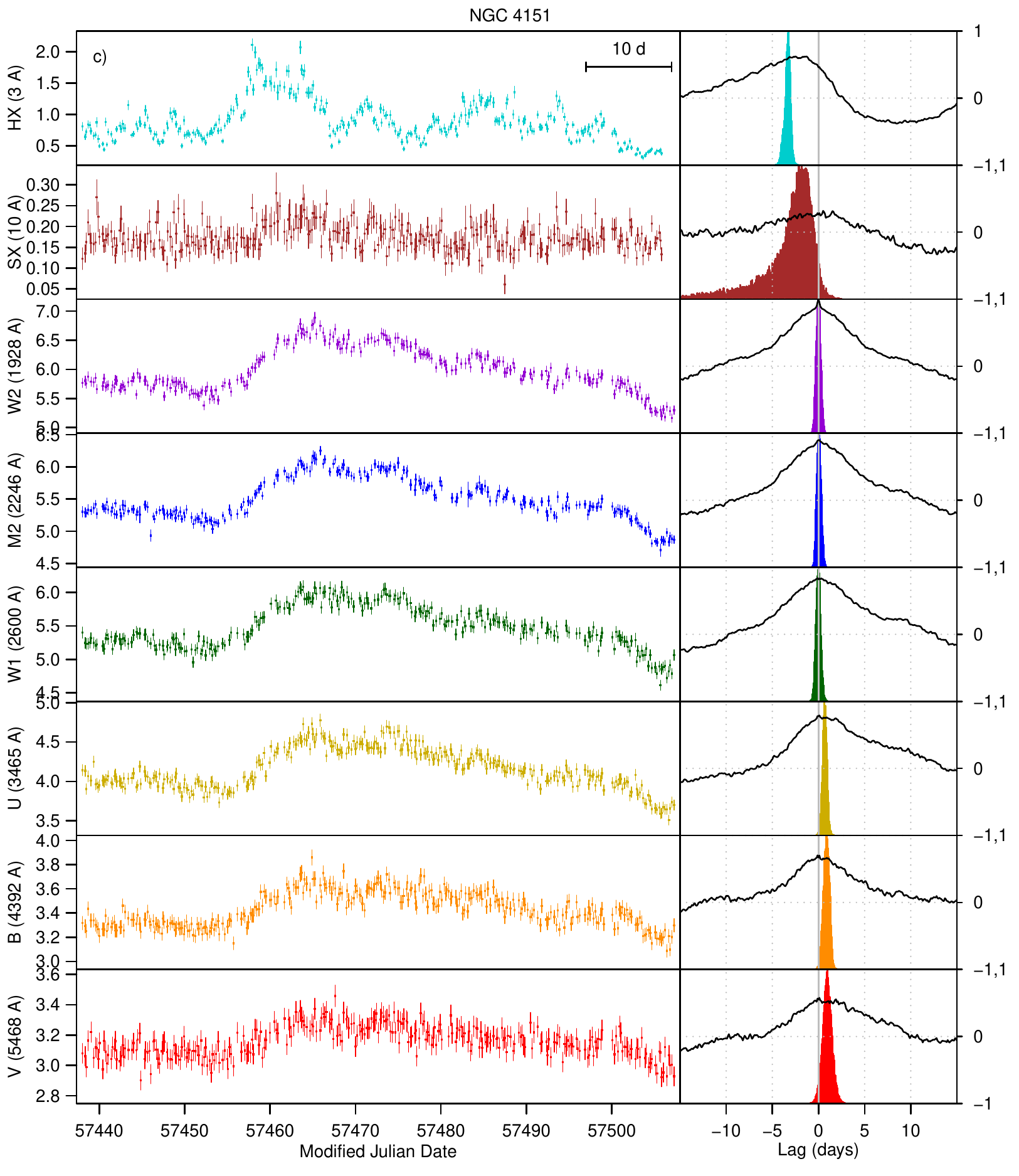} 
\caption{Same as \rb{Figure~1a} except for \ngb.
\ngb\ was the second AGN to be subjected to \swift\ IDRM.
The initial \swift\ analysis is presented in \cite{Edelson17}.
Because \ngb\ is typically the brightest or among the brightest AGN in the sky in most observable wavebands, these light curves are particularly well defined, and the interband lags highly constrained.
It is also the only IDRM target for which \swift\ BAT data could be gathered, extending the coverage into the hard X-rays.}
\label{fig:fig1c}
\end{center}
\end{figure}

\begin{figure}
\figurenum{1d}
\begin{center}
 \includegraphics[width=7in]{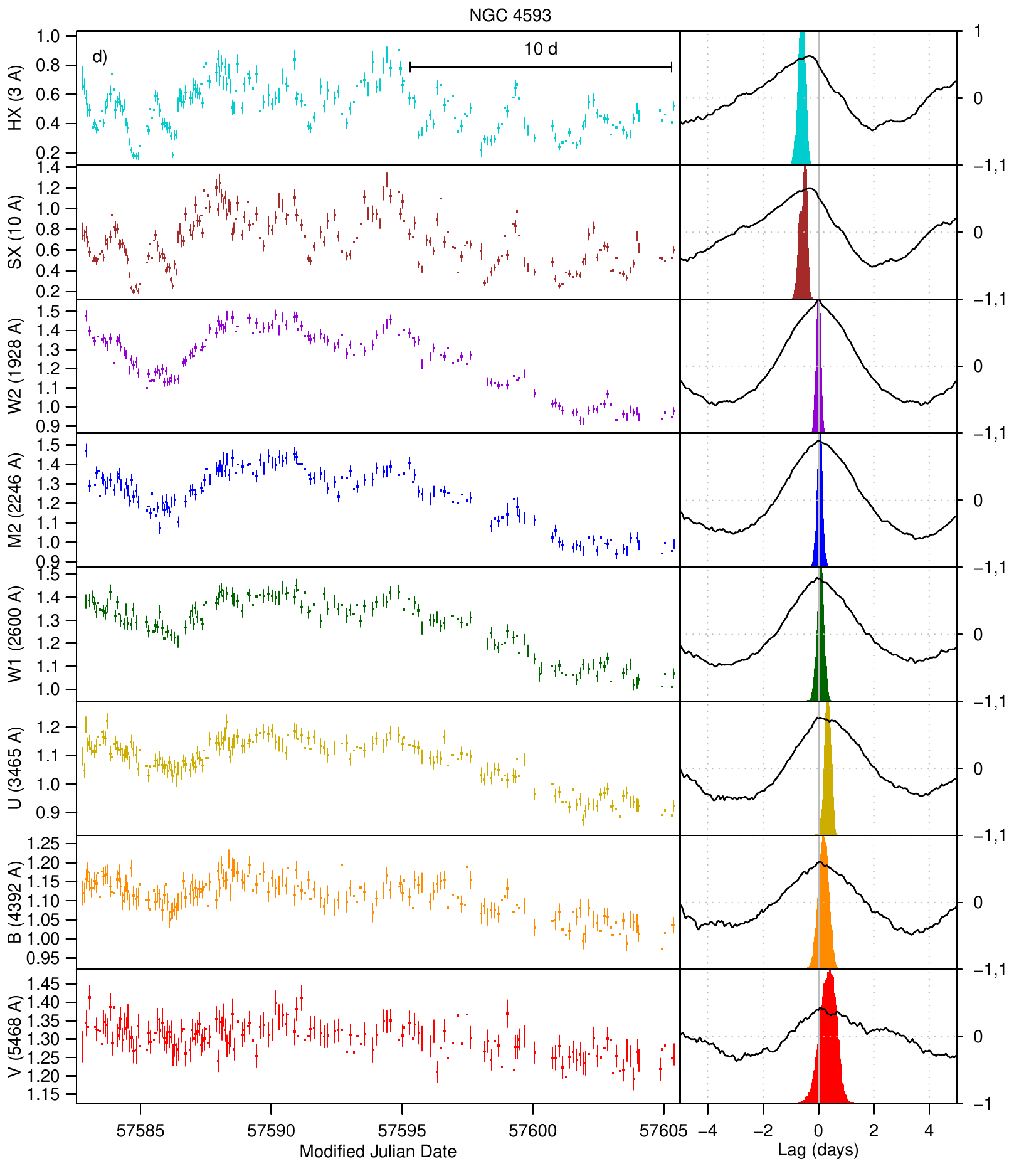} 
\caption{Same as \rb{Figure~1a} except for \ngc.
\ngc\ was the third AGN to be subjected to \swift\ IDRM.
Our initial \swift\ data reduction and analysis is presented in \cite{McHardy18}, and the combined {\it HST}/\swift\ analysis in \cite{Cackett18}.
This AGN has the lowest mass ($\sim 8 \times 10^6 M_\odot $) and highest Eddington ratio ($\sim8$\%) of any IDRM target analyzed to date.}
\label{fig:fig1d}
\end{center}
\end{figure}

\begin{figure}
\figurenum{2}
\begin{center}
 \includegraphics[width=3in]{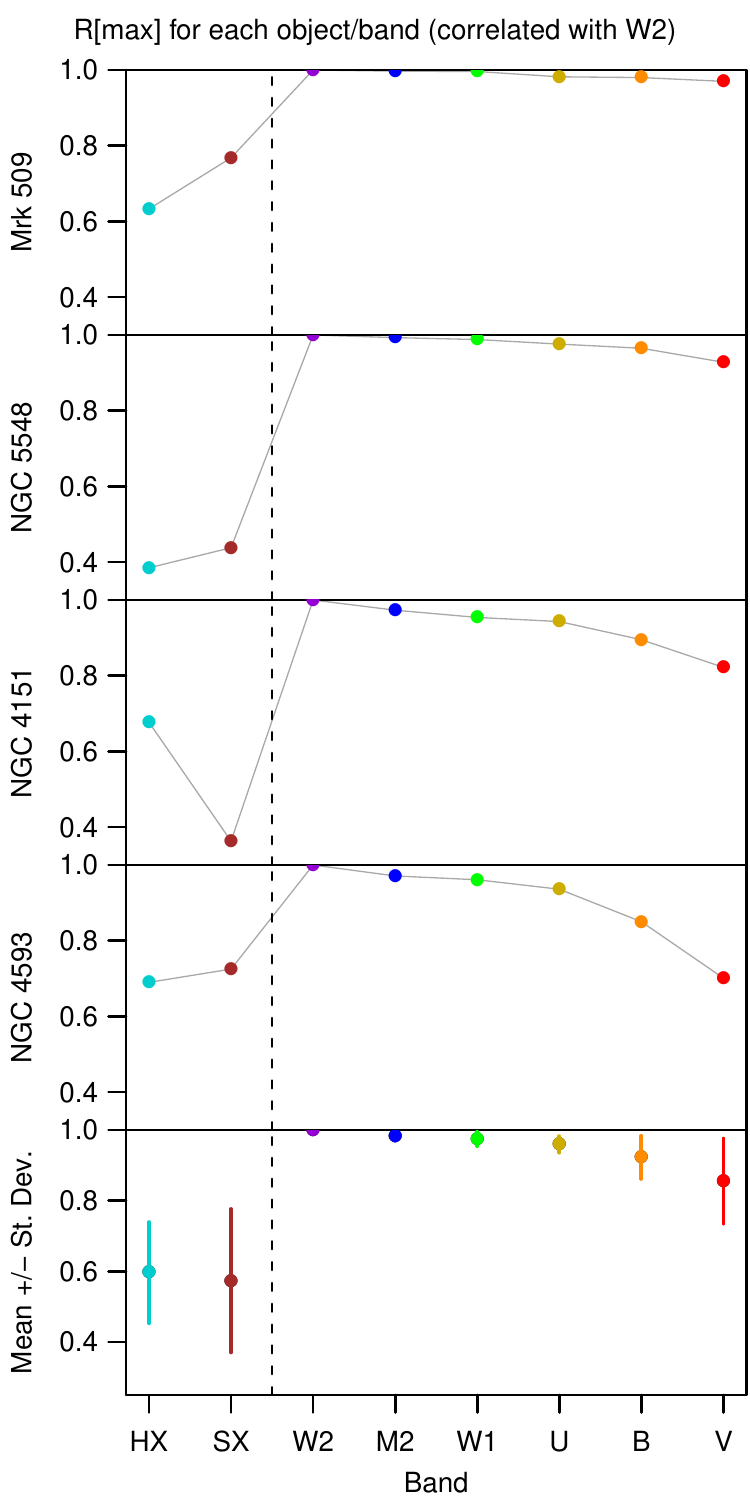} 
\caption{Plots of $r_\mathrm{max}$ for each band with W2, shown as \rb{dots (with colors matching those in the same bands in Figures 1 and 3) with gray lines} to guide the eye.
Each object is shown separately in the top four panels, going from most luminous at the top (\mrk) to least luminous on the bottom (\ngc).
All data are from Column~5 of \rb{Table~3.
For each band the colors are the same as used in Figure~1.
The vertical dashed line shows the separation between the X-ray and UV/optical regimes.}
The bottom panel shows the mean values for each band as black dots, with error bars showing the measured standard deviations.}
\label{fig:fig2}
\end{center}
\end{figure}

\begin{figure}
\figurenum{3}
\begin{center}
 \includegraphics[width=6.5in]{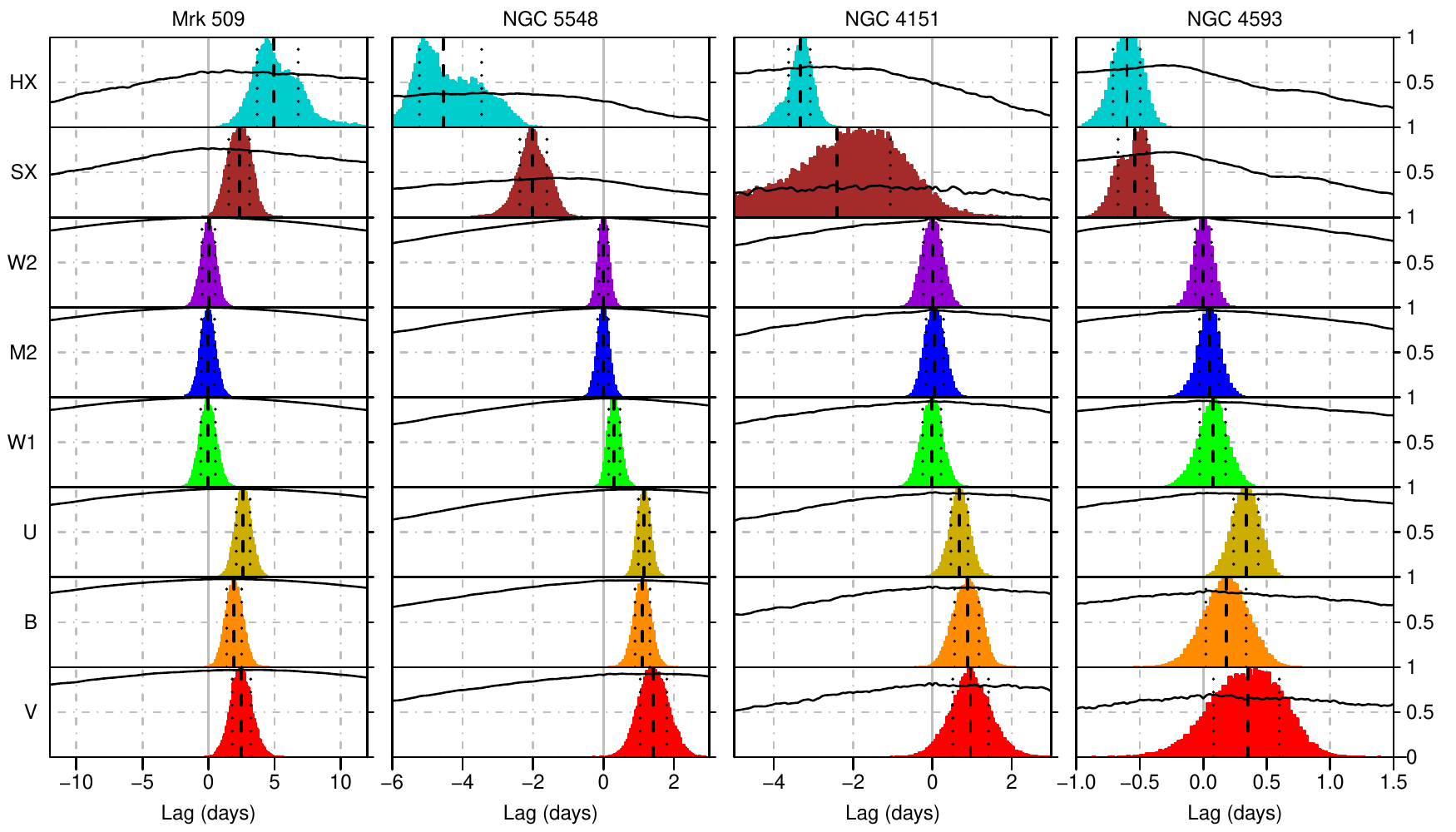} 
 \includegraphics[width=6.5in]{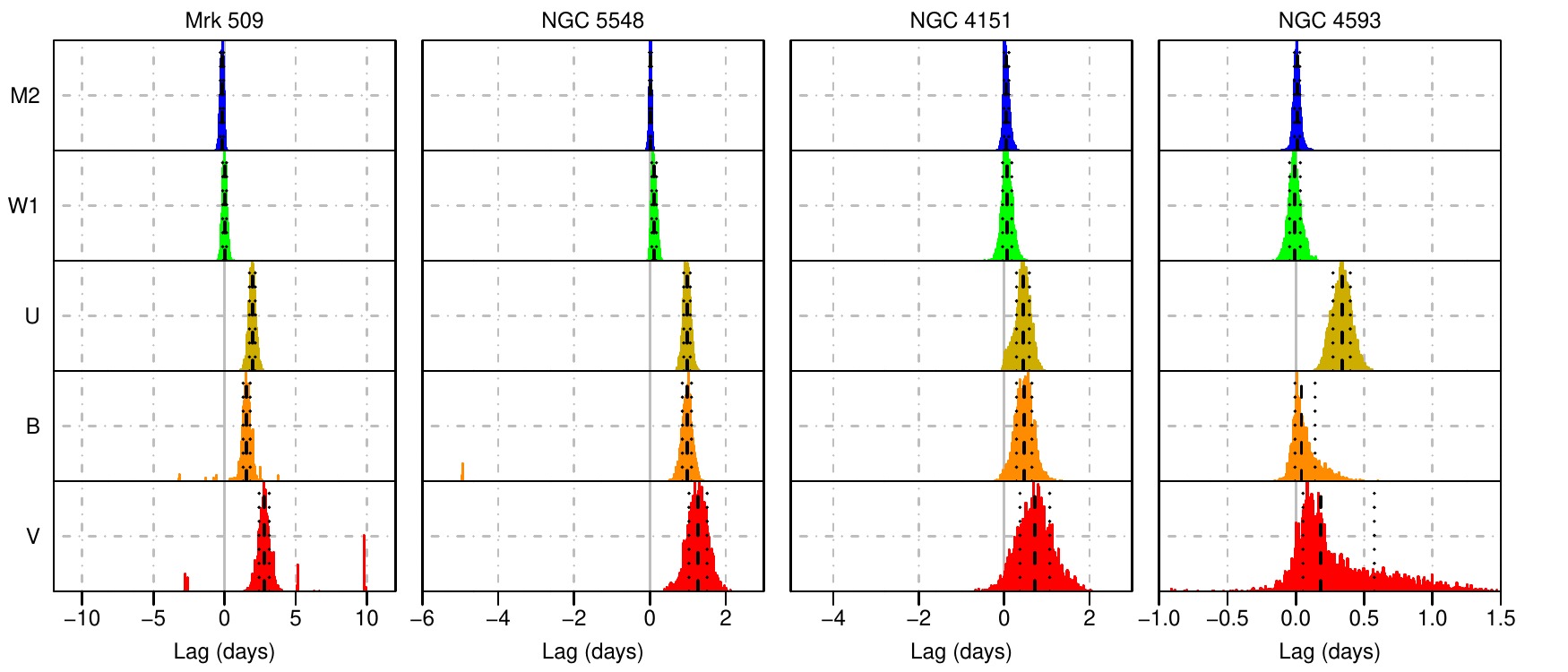} 
\caption{Top: the top set of 32 panels show ICCFs and FR/RSS centroid plots using the same scheme as \rb{Figure~1} (right), except that both {\it x}- and {\it y}-axes have been magnified to focus on the peak.
The dotted horizontal line shows $ r = 0.5 $.
Vertical dashed and dotted black lines indicate the median and the bounds of the 68\% ($\pm 1 \sigma$) confidence intervals derived from the FR/RSS simulations.
Bottom: the bottom set of 20 panels show the \javelin\ simulation results and confidence intervals using the same color scheme and line colors/types.
Because the \javelin\ analysis is restricted to the UVOT data, and the W2 results are identically zero, only results for filters M2, W1, {\it U,} {\it B} and {\it V} are shown.}
\label{fig:fig3}
\end{center}
\end{figure}

\begin{figure}
\figurenum{4}
\begin{center}
 \includegraphics[width=3in]{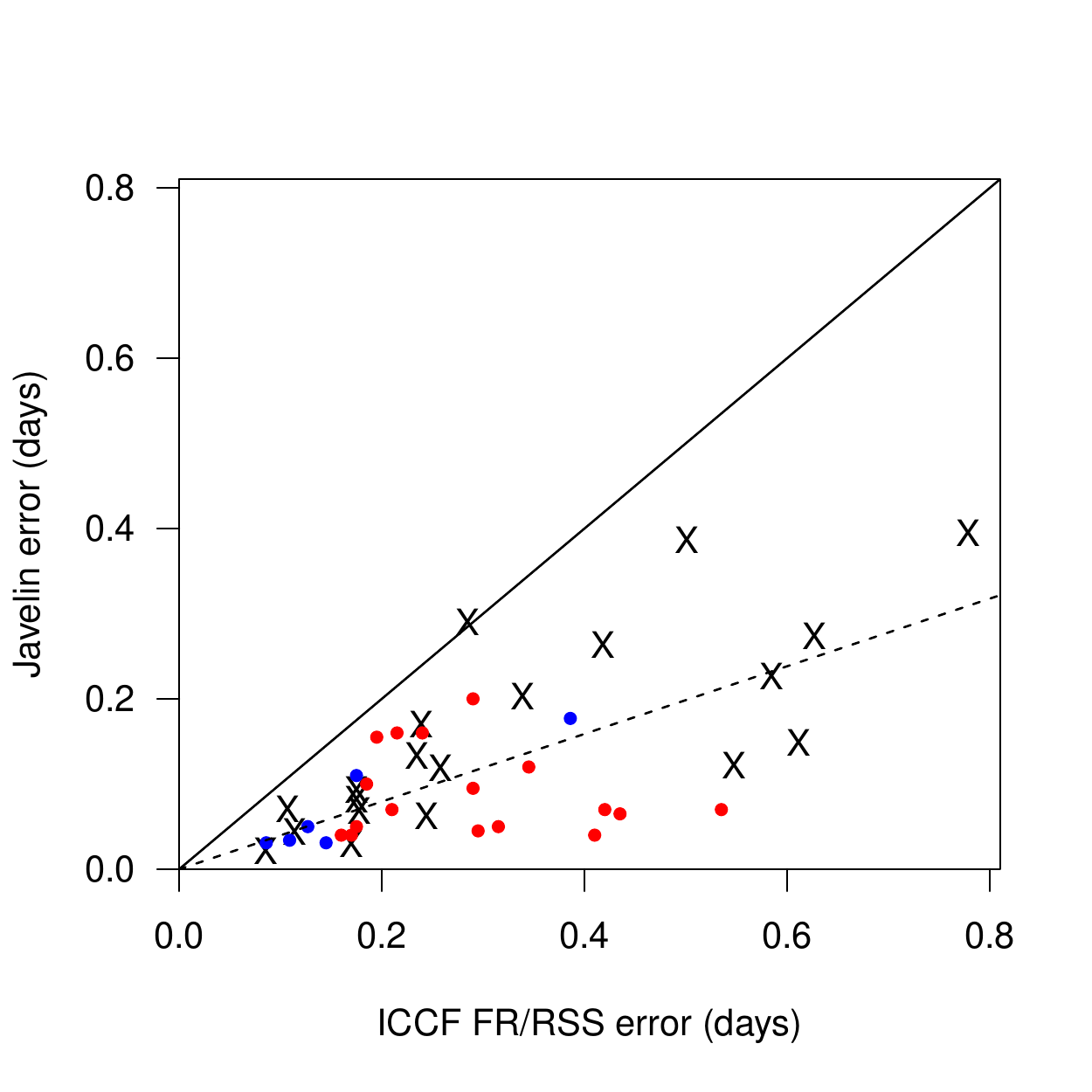} 
\caption{Comparison of ICCF FR/RSS centroid uncertainties and \javelin\ uncertainties for these data (black X's).
Other published IDRM AGN results are shown as dots: \nga\ (\citealt{Fausnaugh16}; in red) and \ngc\ (\citealt{McHardy18}; blue).
The solid line shows where the FR/RSS and \javelin\ uncertainties are equal and the dashed line is a fit to the data forced through the origin $ y = mx $.
Note that all but one of the points lie below the solid line.
The shallow fitted slope, $ m = 0.40 $, indicates that the \javelin\ uncertainties are, on average, a factor of $ 1/0.40 = 2.5 $ smaller than the ICCF FR/RSS uncertainties.
The median ratio of ICCF/\javelin\ lags is similar, 2.6.
The reasons for this large discrepancy are explored in the text.}
\label{fig:fig4}
\end{center}
\end{figure}

\begin{figure}
\figurenum{5}
\begin{center}
 \includegraphics[width=6.5in]{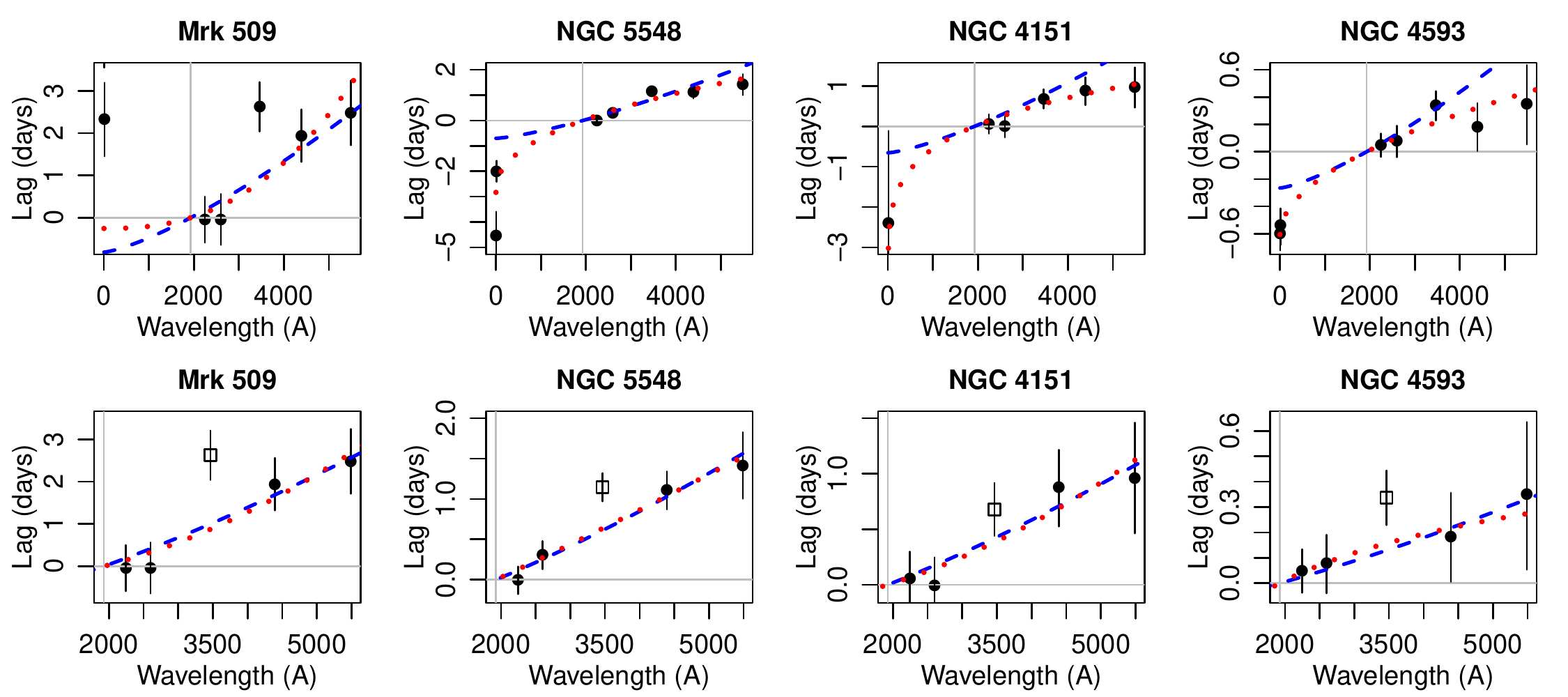} 
\caption{Plots of measured median ICCF centroid lag ($\tau$) as a function of central wavelength ($\lambda$) for all bands in Figure~3.
All lags are measured relative to the W2 band, so that autocorrelation point is not shown.
The red dotted lines show the fit to the function $ \tau = \tau_0\,\left[ (\lambda/\lambda_0)^\alpha - 1 \right] $, where $\tau_0$ is the normalization, $\alpha$ is the power-law index and $\lambda_0$ is the reference band wavelength, 1928~\AA\ for the W2 band.
The gray lines cross at $\lambda=\lambda_0$ because the ACF lag $\tau$ is identically zero.
The blue dashed lines show the same fit but with fixed index $ \alpha = 4/3 $.
The four top panels show the fits for the full data.
In general these functions yield poor fits due to a mismatch in the X-rays, an excess in {\it U} band in all objects, and disagreements in {\it B} and {\it V} in two of the objects.
Thus the bottom four panels show fits restricted just to the UVOT data, excluding {\it U} band.
(The {\it U} band lags are shown as empty boxes because they do not participate in the fit.)
These bottom panels show that once the X-rays and {\it U} band data are excluded, the fits are improved, with acceptable $\chi^2$.
We note that in one source (\ngc) all remaining points are within $\sim 1.2 \sigma $ of zero lag and in another (\ngb) that only two of the remaining points ({\it B} and {\it V}) are significantly above zero.}
\label{fig:fig5}
\end{center}
\end{figure}

\begin{figure}
\figurenum{A1}
\begin{center}
 \includegraphics[width=3.5in]{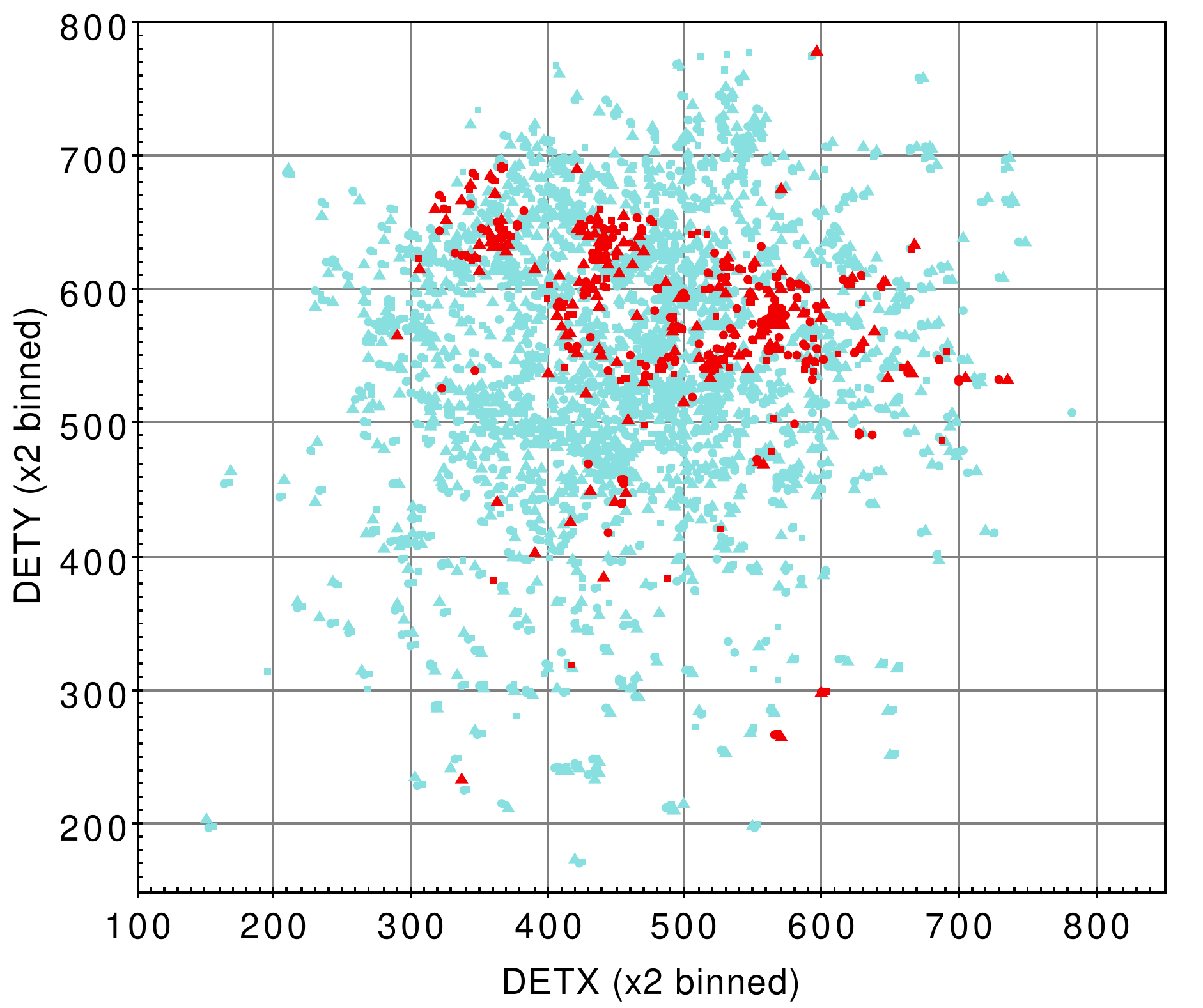} 
 \includegraphics[width=3.5in]{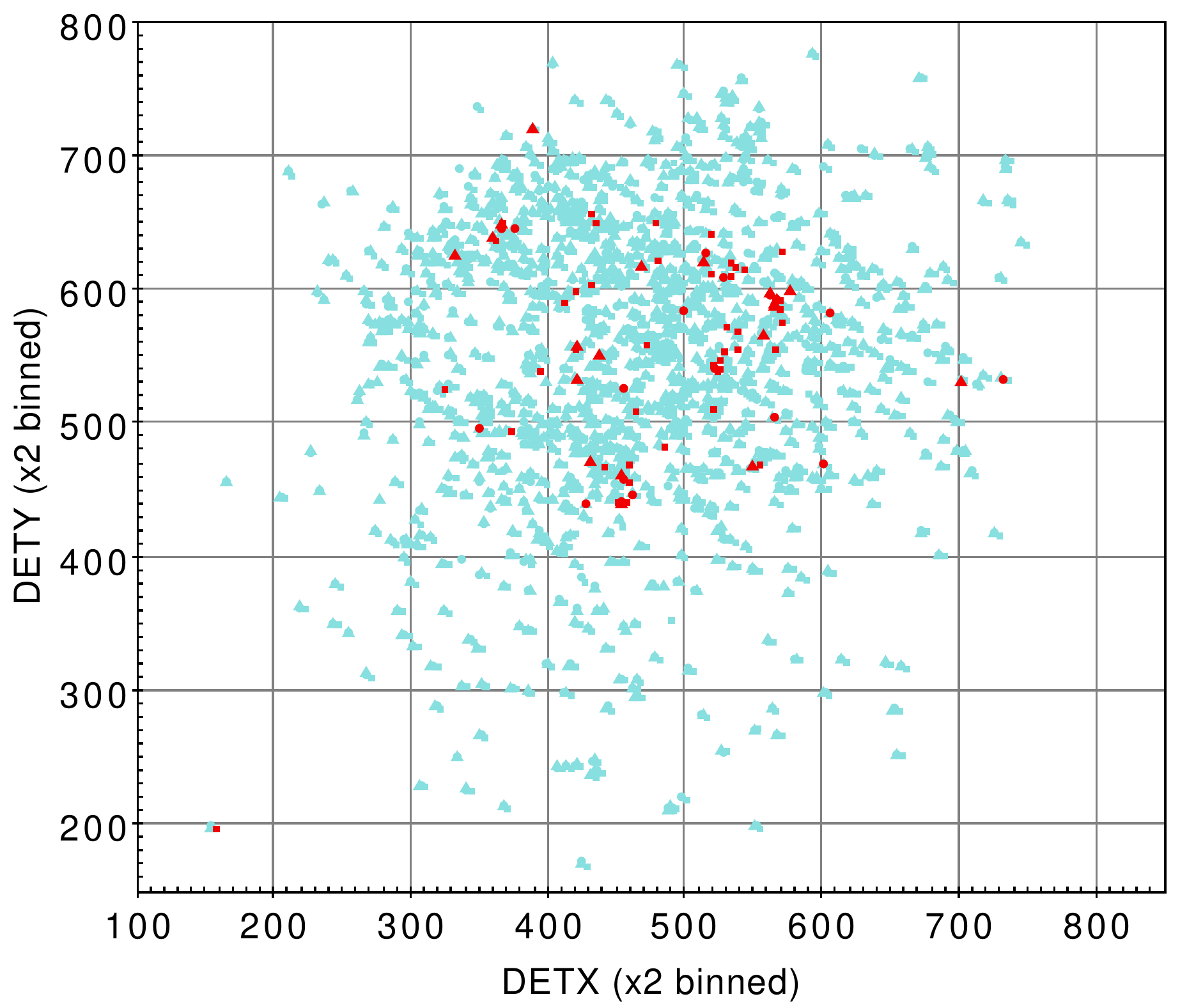} 
\caption{Left: detector plane mapping of all UV data points that were tested for dropouts.  
Point colors reflect dropout status (red = dropout, gray = non-dropout) and shape indicates the UVOT filter (squares for W2, circles for M2, and triangles for W1).
Right: same as above for the optical dropout mapping ({\it U, B} and {\it V} as squares, circles, and triangles, respectively).
Note that the dropouts are highly clustered in both figures, allowing the masking described in this Appendix.
There are far fewer optical dropouts, however, so the optical masking requires fewer boxes than the UV.
The boxes are listed in Tables A.2 and A.3.}
\label{fig:figA1}
\end{center}
\end{figure}

\end{document}